\documentclass[aps,prd,onecolumn,groupedaddress,floatfix,longbibliography,superscriptaddress,notitlepage,10pt]{revtex4-1}

\usepackage{amssymb}
\usepackage[usenames,dvipsnames,svgnames]{xcolor}
\usepackage{subfig}
\usepackage{braket}
\usepackage{slashed}

\usepackage{graphicx}
\usepackage{amsmath}
\usepackage{soul}
\usepackage[normalem]{ulem}

\usepackage{soul}

\usepackage{hyperref}
\hypersetup{colorlinks=true, linkcolor=NavyBlue, citecolor=PineGreen,urlcolor=cyan}

\newcommand{\beq}{\begin{equation}}
\newcommand{\eeq}{\end{equation}}
\newcommand{\barr}{\begin{eqnarray}}
\newcommand{\earr}{\end{eqnarray}}
\newcommand{\mbf}{\mathbf}
\newcommand{\bs}{\boldsymbol}

\begin{document}


\title{Anomalous Hall effect in anisotropic type-II Weyl semimetals}

\author{R. Martínez von Dossow}
\email{ricardo.martinez@correo.nucleares.unam.mx}
\affiliation{Instituto de Ciencias Nucleares, Universidad Nacional Aut\'{o}noma de M\'{e}xico, 04510 Ciudad de M\'{e}xico, M\'{e}xico}

\author{A. Mart\'{i}n-Ruiz}
\email{alberto.martin@nucleares.unam.mx}
\affiliation{Instituto de Ciencias Nucleares, Universidad Nacional Aut\'{o}noma de M\'{e}xico, 04510 Ciudad de M\'{e}xico, M\'{e}xico}

\author{L. F. Urrutia}
\email{urrutia@nucleares.unam.mx}
\affiliation{Instituto de Ciencias Nucleares, Universidad Nacional Aut\'{o}noma de M\'{e}xico, 04510 Ciudad de M\'{e}xico, M\'{e}xico}

\begin{abstract}
We extend our previous analysis [Phys. Rev. D \textbf{109}, 065005 (2024)] of CPT-odd electromagnetic response in tilted, anisotropic Weyl semimetals to the overtilted (type-II) regime, where electron and hole pockets coexist at the Fermi level. Starting from the minimal QED sector of the Standard-Model Extension matched to a lattice-motivated anisotropic Dirac Hamiltonian with tilt, we compute the zero-temperature finite-density effective action nonperturbatively from the vacuum polarization tensor, and corroborate the result using a complementary chiral kinetic theory formulation that consistently incorporates both Fermi-sea and Fermi-surface contributions. In the type-II regime the unbounded linear dispersion necessitates a physical ultraviolet regularization. Implementing a hard momentum cutoff tied to the lattice bandwidth, we show that the CPT-odd, axion-like response remains finite across the type-I to type-II Lifshitz transition, while acquiring tilt- and anisotropy-dependent renormalizations together with nonuniversal, cutoff-sensitive terms governed by the geometry of the electron and hole pockets. As a concrete application, we evaluate the anomalous Hall conductivity in the prototypical type-II Weyl semimetal WTe$_2$, using parameters extracted from first-principles calculations and experiments, and find that Fermi-sea and Fermi-surface contributions are comparable and partially cancel, yielding a finite and strongly anisotropic Hall response characteristic of the overtilted regime.
\end{abstract}

\maketitle 

\section{Introduction}

Topological phases of matter have reshaped condensed-matter physics by showing that global quantum geometry, encoded in Berry phases and topological invariants, stabilizes robust boundary modes and quantized responses largely insensitive to microscopic details and moderate disorder \cite{HasanKane2010,QiZhang2011,Chiu2016RMP}. Symmetry patterns (time-reversal, inversion, and crystalline symmetries) organize the taxonomy of gapped and gapless phases and link transport coefficients to quantum anomalies and effective field theories such as Chern-Simons and axion electrodynamics. Within this broader framework, Weyl semimetals (WSMs) emerge as paradigmatic gapless topological metals.

WSMs are topologically nontrivial conductors where valence and conduction bands touch at isolated Weyl nodes forming local Dirac cones \cite{PhysRevLett.111.027201,Armitage:2017cjs,Yan:2016euz}. By the Nielsen-Ninomiya theorem, nodes appear in opposite-chirality pairs \cite{Nielsen:1983rb} and act as sources/sinks of Berry curvature; thus the WSM phase is protected by a nonzero Berry flux across the Fermi surface. In real crystals, Weyl cones are generally tilted and anisotropic, so their low-energy behavior departs from the Lorentz-invariant ideal. In noncentrosymmetric WSMs, opposite tilts and rotated anisotropies typically occur for each chiral pair. These deviations impact optical and spin-texture properties \cite{10.1038/nmat4457,PhysRevLett.116.096801} and, crucially, anomalous transport. In the type-II (over-tilted) regime, where electron and hole pockets coexist, Fermi-surface contributions become essential for the anomalous Hall response, motivating a compact, cutoff-dependent treatment beyond the ideal cone picture \cite{Soluyanov2015Nature}.

In the ideal type-I limit (isotropic, untilted cone), the WSM's CPT-odd electromagnetic response is described by axion electrodynamics with $\theta ( \mathbf{r} , t ) = 2 ( \mathbf{b} \cdot \mathbf{r} - b _{0} t)$, equivalently encoded by an axion term in the action. Here, $2\mathbf{b}$ and $2b_{0}$ denote the momentum- and energy-separation of the Weyl nodes. This yields the intrinsic anomalous Hall current $\mathbf{j} _{\mbox{\scriptsize Hall}} =( e ^{2} / 2 \pi ^{2}) \, \mathbf{b} \times \mathbf{E}$. The axionic description can acquire higher-derivative corrections when the electromagnetic fields vary on length scales comparable to microscopic ones; such generalized magnetoelectric couplings beyond the leading $\theta\,\mathbf{E}\!\cdot\!\mathbf{B}$ term were derived for three-dimensional topological insulators in Ref.~\cite{sym17040581} {within a quantum field theory (QFT) framework using the derivative expansion method \cite{MartinezvonDossow:2025mxv}}. In our previous work in Ref.~\cite{PhysRevD.109.065005}, we recovered the axionic structure for a tilted, anisotropic (yet type-I) WSM at finite density by matching a fermionic sub-sector  of the  Standard Model Extension (SME) \cite{Colladay1997,Colladay1998} to a linearized lattice Hamiltonian and validating the coefficients via chiral kinetic theory. In that setting, the general type-I CPT-odd (axion-like) coupling takes the form
\begin{equation}
\Theta ( \mathbf{r} , t ) = 2 ( \boldsymbol{\mathcal{B}} \cdot \mathbf{r} - \mathcal{B} _{0} t) ,
\label{typeI-axion-coupling}
\end{equation}
where the effective parameters $\boldsymbol{\mathcal{B}}$ and $\mathcal{B}_{0}$ are tilt- and anisotropy-dependent \cite{PhysRevD.109.065005}. This compactly encodes the axion response in the tilted, anisotropic type-I regime, reduces to the ideal isotropic, untilted result in the appropriate limit ($\boldsymbol{\mathcal{B}} \to \mathbf b$, $\mathcal{B}_0 \to b_0$), and serves as the baseline for the type-II extension developed below.

In this work we extend the analysis to the over-tilted (type-II) regime, where electron- and hole-pocket Fermi surfaces appear and cutoff-dependent finite-density contributions become essential, deriving compact expressions for the Hall response and related CPT-odd tensors. A compact, cutoff-dependent extension to the type-II regime at finite density with generic tilt and anisotropy has been lacking; here we fill this gap by deriving, at $T=0$ and $\mu \neq 0$, a physically motivated hard-cutoff effective action and corroborating it via chiral kinetic theory with a clean Fermi-sea versus Fermi-surface decomposition, obtaining concise formulas for the Hall (CPT-odd) response.

The content of this paper is as follows. Section~\ref{model} introduces the tilted, anisotropic Weyl-semimetal model and fixes our notation. Section~\ref{kinetic_theory_approach} employs chiral kinetic theory to analyze the electromagnetic response in the type-II regime. Section~\ref{field_theory_approach} turns to a field-theoretic treatment: we map the WSM model onto a fermionic sub-sector of the SME, choose parameters that encode tilt and anisotropy, and derive the CPT-odd effective action via the vacuum polarization tensor. As expected, both approaches yield the same axion-like structure. The explicit relation between the SME coefficients and the condensed-matter parameters is derived in Appendix~\ref{APPA}. In Section~\ref{transport_coefficient} we evaluate the resulting integrals for the transport coefficient (i.e. the conductivity), clearly separating Fermi-sea and Fermi-surface contributions, with the corresponding integration domains detailed in Appendices~\ref{Angular_domain_app} and~\ref{app:boundary_selector_new}. Section~\ref{Applications} applies the formalism to the type-II material WTe$_2$, extracting compact expressions for the anomalous Hall response. Section~\ref{conclusion} summarizes our findings and presents the conclusions.

\section{Tilted-anisotropic Weyl nodes: Model and notation} \label{model}

We consider a minimal two-node Weyl semimetal with opposite chiralities $\chi = \pm 1$, separated in momentum and energy, and neglect nonuniversal band bending away from the nodes so that a linear description holds within a finite window. The low-energy Hamiltonian around node $\chi$ is
\begin{align}
H_{\chi} (\mbf{k}) = {\mbf v}_{\chi} \cdot ({\mbf k} +  \chi {\mbf{\tilde b}})
-  \chi {\tilde b}_{0} \sigma _{0} + \chi  
({\mbf k} + \chi {\mbf {\tilde b}}) 
\mathbb{A}_{\chi} \boldsymbol{\sigma},
\qquad {\mbf W} 
{\mathbb A}_\chi \bs{\sigma} \equiv W^{i} A _{\chi\, ij } \bs{\sigma}^j ,    \label{Hamiltonian_Tilting}  
\end{align} 
where $\mathbf{k}$ is the crystal momentum, $\sigma_{0}$ the identity matrix, and $\boldsymbol{\sigma}=(\sigma_{x},\sigma_{y},\sigma_{z})$ the vector of Pauli matrices. The matrix $\mathbb{A}_{\chi}=[A_{\chi,ij}]$ encodes the anisotropic Fermi velocities, while $\mathbf{v}_{\chi}$ is the tilt velocity. The terms $-\chi\,\widetilde{\mathbf b}$ and $-\chi\,\widetilde b_{0}$ fix the location of the node of chirality $\chi$ in momentum (relative to $\mathbf{k}=\mathbf{0}$) and in energy (relative to the zero-energy plane), so the two nodes are separated by $2\widetilde{\mathbf b}$ in momentum and by $2\widetilde b_{0}$ in energy.

Introducing the shifted momentum $ {\boldsymbol K } _{\chi} =  {\mbf k} + \chi {\mbf {\tilde b}} $, the anisotropy-rescaled momentum $\mathcal{K} _{\chi i} =  A _{\chi ij } K _{\chi j} $, and the dimensionless (rescaled) tilt $\mathcal{V} _{\chi i} =  A_{\chi ij} ^{-1} \mathrm{v} _{\chi j}$, the Hamiltonian becomes
\begin{align}
H _{\chi} (\mbf{k}) = \mathcal{V} _{\chi i}  \mathcal{K} _{\chi i} -   \chi {\tilde b}_{0} \sigma _{0} + \chi  \mathcal{K} _{i} \sigma _{i}   .   
\end{align}
The band dispersion and group velocity read
\begin{align}
E _{s\chi}(\mathbf{k}) &= -   \chi {\tilde b}_{0}  + \mathcal{V} _{\chi i}  \mathcal{K} _{\chi i} + s \, \mathcal{K} _{\chi} , \label{Energy} \\[5pt] v _{s \chi i}(\mathbf{k}) &= \frac{\partial E _{s\chi}}{\partial k ^{i}} = A _{\chi ij} \left( \mathcal V _{\chi j} + s \, \frac{\mathcal K _{\chi j}}{ \mathcal{K} _{\chi} } \right) , \label{Velocity_Tilting}
\end{align}
where $s = \pm 1$ is the band index and $\mathcal{K} _{\chi} \equiv \|\boldsymbol{\mathcal K}_{\chi}\| =  \sqrt{ \mathcal{K} _{\chi i} \mathcal{K} _{\chi i} }$.

In the $\boldsymbol{\mathcal K} _{\chi}$ variables all directional dependence enters through $\boldsymbol{\mathcal V} _{\chi}$. This makes the type-I/type-II classification transparent:
\begin{align}
    \|\boldsymbol{\mathcal V}_{\chi}\|<1 \ \Rightarrow\ \text{type-I} , \qquad \|\boldsymbol{\mathcal V} _{\chi}\| = 1 \ \Rightarrow\ \text{critical (Lifshitz)} , \qquad \|\boldsymbol{\mathcal V} _{\chi}\| > 1  \ \Rightarrow\ \text{type-II}.
\end{align}
The effect of tilt on the Fermi surface follows directly from the dispersion. Setting $E _{s\chi}(\mathbf{k}) = \mu$ and writing $\boldsymbol{\mathcal K} _{\chi} = \mathcal K _{\chi} \,\hat{\mathbf n}_{\mathcal K}$ gives
\begin{align}
    \mu + \chi \, \widetilde b _{0} = \mathcal K _{\chi} \big( s +  \boldsymbol{\mathcal V} _{\chi} \cdot \hat{\mathbf n} _{\mathcal K} \big) ,
\end{align}
so along a fixed direction $\hat{\mathbf n}_{\mathcal K}$ the radial solution is
\begin{align}
    \mathcal K _{s \chi} (\hat{\mathbf n} _{\mathcal K} ; \mu ) = \frac{ \mu + \chi \widetilde b _0}{ s + \boldsymbol{\mathcal V} _{\chi} \cdot \hat{\mathbf n} _{\mathcal K} }.
\end{align}
In the type-I case, the denominators never vanish, so the constant-energy surfaces near the Weyl point are closed. For type-II nodes, however, certain directions fulfill $|\boldsymbol{\mathcal V}_{\chi} \cdot \hat{\mathbf n} _{\mathcal K}| > 1$ such that, for small but finite $| \mu + \chi \, \widetilde b _0 |$, the conduction and valence bands intersect, leading to simultaneous electron and hole pockets and the emergence of open Fermi surfaces.

Exactly at the node energy ($\mu = - \chi \, \widetilde b_{0}$), the linearized model predicts $\boldsymbol{\mathcal V}_{\chi} \cdot \hat{\mathbf n}_{\mathcal K} = -s$, marking the boundary of validity of the linear approximation. In the type-II case, weak quadratic (lattice) corrections become relevant and smooth this crossing into finite electron and hole pockets, while for $|\boldsymbol{\mathcal K}_{\chi}| \lesssim \Lambda$ the linear theory remains reliable. The parameter $\Lambda$ denotes the physical hard momentum cutoff delimiting the linearization window around each node, i.e., the scale at which quadratic band-curvature terms become comparable to the linear ones. In a tight-binding crystal with lattice spacing $a$, one expects $\Lambda \sim \mathcal{O}(\pi/a)$ (up to material-dependent numerical factors), and because of anisotropy, the cutoff surface is ellipsoidal in the original $\mathbf{k}$ coordinates: $|\boldsymbol{\mathcal K}_{\chi}| = |\mathbb{A}_{\chi}\boldsymbol{K} _{\chi}| \le \Lambda$, equivalently $|\mathbb{A}_{\chi}\mathbf{k}| \lesssim \Lambda$ around each node. Within this window ($|\boldsymbol{\mathcal K}_{\chi}| \lesssim \Lambda$), the linear theory provides an accurate description and will be used to define a cutoff-dependent effective action below.

\begin{figure}
    \centering
    \includegraphics[width=0.7\linewidth]{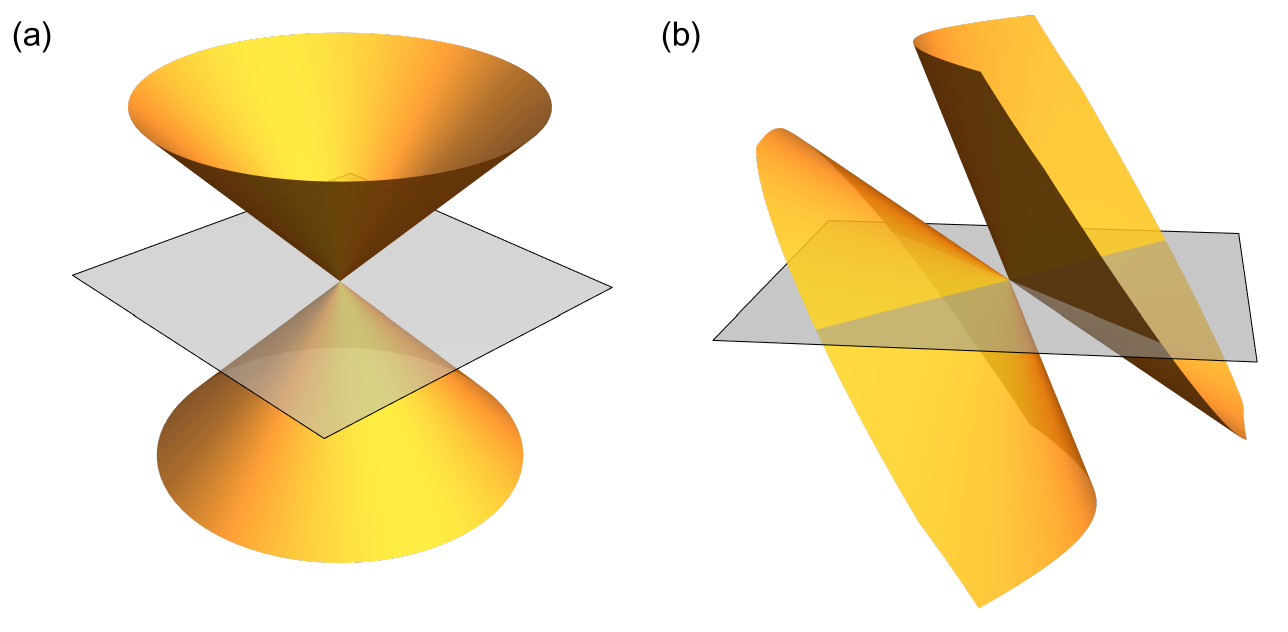}
    \caption{ Type-I versus type-II Weyl cones. The gray plane denotes zero energy.  In type-I (a) the plane intersects the spectrum only at the Weyl point, whereas in type-II (b) it cuts the two branches, explicitly showing the coexistence of electron- and hole-like states.  This feature is responsible for the nonvanishing Fermi-sea contribution and for the necessity of summing over both bands in the anomalous Hall response.
}    \label{cones_fig}
\end{figure}

In the isotropic limit with tilt, $\mathbb{A} _{\chi} = v _{F}  \mathbb{I}$ and thus $\boldsymbol{\mathcal K} _{\chi} = v _{F}  {\boldsymbol K } _{\chi}$ while $\boldsymbol{\mathcal V} _{\chi} = \mathbf{v} _{\chi} / v _{F}$. The dispersion simplifies to $E _{s \chi}(\mathbf{k}) = - \chi \, \widetilde b _{0} + \mathbf{v} _{\chi} \cdot {\boldsymbol K } _{\chi} + s \, v _{F} \, \| {\boldsymbol K } _{\chi} \| $, and the type-I/II criterion reduces to $|\mathbf{v}_{\chi}|/v_{F}\lessgtr 1$, with the critical (Lifshitz) point at $\| \mathbf{v}_{\chi} \| = v _{F}$. The ``most over-tilted'' direction is simply $\hat{\mathbf n} ^{*} \parallel \mathbf{v} _{\chi}$, and the Fermi surface evolves from closed spheres (type-I) to open sheets with coexisting electron- and hole-pockets (type-II). This geometric distinction is illustrated in Fig.~\ref{cones_fig}, where the zero-energy plane intersects the spectrum only at the node in the type-I case, while cutting both bands in the overtilted type-II regime.

We close this section by recalling the two complementary routes we use to obtain the electromagnetic response of a generic WSM under external fields: (i) the chiral kinetic theory framework \cite{PhysRevLett.109.162001} and (ii) the one-loop effective action of quantum electrodynamics \cite{Dittrich:2000zu,Dittrich:1985yb}. In Sections~\ref{kinetic_theory_approach} and \ref{field_theory_approach} we briefly review both derivations for the general tilted, anisotropic model introduced above without yet distinguishing between type-I and type-II; that distinction will enter only at the level of the transport coefficients (e.g. the anomalous Hall conductivity) evaluated later. Since these developments were already presented in our previous work \cite{PhysRevD.109.065005}, the material here is a concise recap provided for completeness and to fix notation. 

\section{Berry-curvature semiclassics: The chiral kinetic route} \label{kinetic_theory_approach}

Chiral kinetic theory treats the many-electron crystal as a continuum of wave packets governed by a Boltzmann equation whose dynamics and phase-space measure are modified by the Berry curvature. Rather than reproducing the semiclassical equations of motion, we refer to standard derivations and modern expositions \cite{SundaramNiu1999, RevModPhys.82.1959, PhysRevB.108.155132, Medel2}. The general expression for the Berry curvature-induced current in the presence of an electric field ${\bf{E}}$ is
\begin{align}
{\bf{J}}  &= - \frac{e ^{2}}{\hbar} {\bf{E}} \times \sum _{ s = \pm 1} \sum _{ \chi = \pm 1} \int \frac{d ^{3} {\bf{k}} }{(2 \pi ) ^{3}} \,  \boldsymbol{\Omega} _{s \chi} ({\bf{k}}) \,  f _{s \chi} ^{\mbox{\scriptsize F.D.}} ({\bf{k}}) ,  \label{CurrentDensity}
\end{align}
where $f^{\mathrm{FD}}_{s\chi} ({\bf{k}})$ is the Fermi-Dirac distribution for Bloch electrons with chirality $\chi$ in the $s$-th band and $\boldsymbol{\Omega} _{s \chi } ({\bf{k}}) = i \bra{  \nabla _{{\bf{k}}}  u _{s \chi} ({\bf{k}}) } \times  \ket{  \nabla _{{\bf{k}}} u _{s \chi} ({\bf{k}}) } $ is the Berry curvature. The Bloch states $\ket{u _{s \chi} ({\bf{k}}) } $ are defined by $\hat{H} _{\chi} \ket{u _{s \chi} ({\bf{k}}) } = E _{s \chi} \ket{u _{s \chi} ({\bf{k}}) }$, where $\hat{H} _{\chi}$ is the single particle Hamiltonian for a Weyl fermion with chirality $\chi$ given by Eq. (\ref{Hamiltonian_Tilting}).

In Weyl semimetals, the semiclassical description is valid as long as the applied fields are weak and slowly varying, ensuring that carrier dynamics remains confined to the linear regime around each node. We denote by $\Lambda$ the characteristic energy-momentum scale delimiting this regime, beyond which lattice or band-curvature corrections become relevant. Accordingly, the conditions $\hbar\omega,\ \hbar v_{F}|{\bf q}|\ll \Lambda $ must hold, and the field amplitudes must be sufficiently small to keep the electronic motion within the linearized region of the spectrum.

For the present model it is convenient to work in the anisotropy-rescaled momentum $\boldsymbol{\mathcal K}_{\chi}= \mathbb A_{\chi}\mathbf {\boldsymbol K } _{\chi}$, where $\mathbb A_{\chi}$ collects the Fermi-velocity coefficients. In these variables, the Berry curvature of band $s = \pm 1$ at node $\chi = \pm 1$ takes the form
\begin{equation}
\Omega_{s\chi i}(\boldsymbol{\mathcal K}_{\chi})
= -\,\frac{s\chi}{4}\,\epsilon_{ijk}\,\epsilon_{lsr}\,
A_{\chi jl} A_{\chi ks}\,
\frac{(\mathcal K_{\chi})_{r}}{\mathcal K_{\chi}^{3}},
\end{equation}
which can be simplified using the tensor identity
$\epsilon_{pqr}\,\det(\mathbb A_{\chi})\,(A_{\chi}^{-1})_{rk}
= \epsilon_{ijk}\,A_{\chi ip}A_{\chi jq}$,
to yield
\begin{equation}
\Omega_{s\chi i}(\boldsymbol{\mathcal K}_{\chi})
= -\,\frac{s\chi}{2}\,\det(\mathbb A_{\chi})\,(A_{\chi}^{-1})_{ri}\,
\frac{(\mathcal K_{\chi})_{r}}{\mathcal K_{\chi}^{3}} .
\label{BerryCurvature}
\end{equation}
Two brief comments are useful here. First, the tilt $\mathbf v _{\chi}$ does not enter $\boldsymbol{\Omega} _{s\chi}$ because it multiplies $\sigma _{0}$ and leaves the eigenvectors unchanged; the curvature is fully controlled by the anisotropy matrix $\mathbb{A} _{\chi}$ and the direction of $\boldsymbol{\mathcal K} _{\chi}$. Second, \eqref{BerryCurvature} describes a monopole field centered at the node: it is odd under $\boldsymbol{\mathcal K} _{\chi} \to - \boldsymbol{\mathcal K} _{\chi}$, decays as $1/\mathcal {K} _{\chi}^{2}$, and its flux through any closed surface enclosing the node yields the quantized charge $\pm 2 \pi$, with the overall sign fixed by $s \chi \, \mathrm{sgn} ( \det\mathbb{A} _{\chi} )$. In the isotropic limit $\mathbb{A} _{\chi} = v _{F} \mathbb{I}$ one recovers the familiar result $\boldsymbol{\Omega} _{s\chi} = - (s \chi / 2 ) \, \boldsymbol{\mathcal{K}} _{\chi} / \mathcal{K} _{\chi} ^{3}$.

From Eqs.~\eqref{BerryCurvature} and \eqref{Energy} it is natural to evaluate the current in the rescaled variables $\boldsymbol{\mathcal{K}} _{\chi} = \mathbb{A} _{\chi} {\boldsymbol{K}} _{\chi}$, because the cone becomes isotropic in $\boldsymbol{\mathcal{K}} _{\chi}$, the occupation $f ^{\mathrm{FD}} _{s \chi}$ depends on $\boldsymbol{\mathcal{K}} _{\chi}$ only through $-\chi \, \widetilde b _{0} + \boldsymbol{\mathcal{V}} _{\chi} \cdot \boldsymbol{\mathcal{K}} _{\chi} + s \, \| \boldsymbol{\mathcal{K}} _{\chi} \|$, and the cutoff is simply $\| \boldsymbol{\mathcal{K}} _{\chi} \| \lesssim \Lambda$. We first shift to the node ${\boldsymbol{K}}  _{\chi} = \mathbf{k} + \chi \, \widetilde{\mathbf{b}}$, which leaves the measure invariant, $d ^{3} \mathbf{k} = d ^{3} {\boldsymbol{K}} _{\chi}$. We then perform the anisotropy rescaling $\boldsymbol{\mathcal{K}} _{\chi} = \mathbb{A} _{\chi} {\boldsymbol{K}} _{\chi}$, whose Jacobian is constant, $d ^{3} {\boldsymbol{K} } _{\chi} = \det( \mathbb{A} _{\chi} ^{-1}) \, d ^{3} \boldsymbol{\mathcal{K}} _{\chi}$. With these steps, Eq.~\eqref{CurrentDensity} becomes
\begin{align}
{\bf{J}}  &= - \frac{e ^{2}}{\hbar} {\bf{E}} \times 
\sum_{s=\pm1}\sum_{\chi=\pm1}
\det(\mathbb A_{\chi}^{-1})
\int \frac{d^{3}\boldsymbol{\mathcal K}_{\chi}}{(2\pi)^{3}}\,
\boldsymbol{\Omega} _{s\chi }(\boldsymbol{\mathcal K}_{\chi})\,
f^{\mathrm{FD}}_{s\chi}(\boldsymbol{\mathcal K}_{\chi}).
\label{eq:J_after_change}
\end{align}
Inserting the Berry curvature \eqref{BerryCurvature} and using $\det ( \mathbb{A} _{\chi} ^{-1}) \det ( \mathbb{A} _{\chi} ) = 1 $ yields
\begin{align}
{\bf{J}}
= \frac{e ^{2}}{2 \hbar} {\bf{E}} \times 
\sum_{s=\pm1}\sum_{\chi=\pm1} s\,\chi\,
\int \frac{d^{3}\boldsymbol{\mathcal K}_{\chi}}{(2\pi)^{3}}\,
\frac{ ( \mathbb{A} _{\chi} ^{-1} ) ^{T} \, \boldsymbol{\mathcal K}_{\chi} }{\mathcal K_{\chi}^{3}}\,
f^{\mathrm{FD}}_{s\chi}(\boldsymbol{\mathcal K}_{\chi}),
\label{eq:J_final_rescaled}
\end{align}
so that all anisotropy is carried explicitly by the matrix $\mathbb{A} _{\chi} ^{-1}$.

In the $T \to 0$ limit at finite chemical potential $\mu$, the Fermi-Dirac occupation $f ^{\mathrm{FD}}_{s\chi}$ reduces to a Heaviside step function $H(x)$. The basic integral entering the current therefore reads
\begin{equation}
\label{Integral}
\boldsymbol{\mathcal{I}} _{\chi s} = \int \frac{d ^{3} \boldsymbol{\mathcal{K}} _{\chi}}{( 2 \pi ) ^{3}} \; \frac{   \boldsymbol{\mathcal{K}} _{\chi}  }{ \| \boldsymbol{\mathcal{K}} _{\chi} \| ^{3}} \; H \big[ \mu - { E _{s\chi}(\mathbf{k}) } \big] ,
\end{equation}
where $E _{s\chi}(\mathbf{k})$ is the band dispersion. The current density then becomes
\begin{equation}
\label{CurrentDensityFinal}
{\bf{J}}
= \frac{e ^{2}}{2 \hbar} {\bf{E}} \times  
\sum_{s=\pm1}\sum_{\chi=\pm1}
s \, \chi   \left[ ( \mathbb{A} _{\chi} ^{-1} ) ^{T}\, \boldsymbol{\mathcal I}_{\chi s} \right] ,
\end{equation}
where $\boldsymbol{\mathcal I}_{\chi s} $ is given by Eq.~\eqref{Integral}.

This compact form will be our starting point for the evaluation of the current and, subsequently, the anomalous Hall conductivity. At $T=0$, the step function in Eq.~\eqref{Integral} induces a natural split into Fermi-sea and Fermi-surface pieces: the former reproduces the axion (“sea’’) response, while the latter encodes the pocket contributions that become essential in the type-II regime. Since the integrand depends on $\boldsymbol{\mathcal{K}} _{\chi}$ only through its direction and magnitude, we thus perform the angular integration on the sphere, reduce the problem to a radial integral within the linear window $\| \boldsymbol{\mathcal{K}} _{\chi} \| \lesssim \Lambda$, and track how tilt and anisotropy enter solely via the boundary condition $H \big[ \mu - {E _{s\chi}(\mathbf{k}) } \big]$ and the tensor factor $(A _{\chi} ^{-1}) _{ir}$. In this way we show that the axionic structure is retained, with coefficients renormalized by tilt and anisotropy.

\section{The effective field theory approach} \label{field_theory_approach}

From the perspective of fundamental interactions in Lorentz Invariance Violation (LIV) QED, the initial step for any calculation is the interacting electron-photon action given by the corresponding fermionic sector of the SME
\begin{eqnarray}
&& S = \int d ^{4} x \, \bar{\Psi} \left( \Gamma ^{\mu} i \partial _{\mu} - M
- e \Gamma ^{\mu} A _{\mu} \right) \Psi, \label{BASICGENACT}  \\
&&\Gamma ^{\mu }={ \gamma^\mu}+c^{\mu }{}_{\nu }\gamma ^{\nu }+d^{\mu
}{}_{\nu }\gamma ^{5} \gamma ^{\nu }+ e^\mu + if_\mu \gamma^5+ \frac{1}{2} g^{\kappa \lambda \mu} \sigma_{\kappa \lambda },\\
&& M=m +im_5 \gamma^5 + a_{\mu }\gamma ^{\mu }+b_{\mu
}\gamma ^{5} \gamma ^{\mu } + \frac{1}{2}H^{\mu\nu} \sigma_{\mu \nu}.
\end{eqnarray}
These coefficients represent the most general dimension-three and dimension-four bilinear fermionic operators, which correspond to renormalizable interactions and are constructed from various gamma matrix combinations and their associated tensorial counterparts with corresponding dimensions. Here $\gamma^\mu$ and $\gamma^5$ are the standard Dirac matrices and  $A_\mu$ is the electromagnetic potential.  Crucially, LIV effects within this action must be uniquely (non-redundantly) specified, so it is mandatory to determine which of  the coefficients  (or combinations of them) would  never  show up in an observable  from  a given experimental setting performed in a concordant frame. 
Since  the action (\ref{BASICGENACT}) is considered as the LIV extension of the basic fermionic QED action
\beq
S = \int d ^{4} x \, \bar{\Psi} \left( \gamma ^{\mu} i \partial _{\mu} - m
- e \gamma ^{\mu} A _{\mu} \right) \Psi,
\label{STANDARD}
\eeq 
it may happen that a field redefinition $\Psi \,\,  \rightarrow \,\,  \Theta $ with $\Psi= \mathbb{A} \,  \Theta$, ( $\mathbb{A}$ being an invertible operator), could yield terms of the kind included in $\Gamma^\mu$ and $M$ when the new action is written in terms of $\Theta$.
Since these redefinitions are effectively invertible maps that describe identical physics, the apparently LIV terms induced in this way are not physically relevant and must  be discarded. Subsequently, the calculation will  follow standard QED procedures, with the action (\ref{STANDARD}) augmented only  by the genuinely identified LIV interactions.

The challenge of field redefinition in the Standard Model Extension (SME) emerged with its initial formulation \cite{Colladay1997} and has since been a recurring subject in research \cite{Colladay2002,Lehnert2004,Altschul2006,Lehnert2006,Kosteleck2010, Kosteleck2022}.  
Given that LIV in fundamental interactions are anticipated  to be extremely suppressed, field redefinitions are predominantly examined to the first order in the relevant parameters. Ref. \cite{Altschul2006} stands out as an exception, conducting a nonperturbative analysis. For a definitive simplification of the general fermion action Eq. (\ref{BASICGENACT}), which isolates only the physically significant parameters by removing redundant coefficients, Ref. \cite{Colladay2002} offers a comprehensive approach. This reduced action defines the precise fermionic framework necessary for calculating processes within LIV QED, conceptualized as a fundamental theory.

However, incorporating the LIV fermionic contributions from the SME requires a change in perspective when calculating effective electromagnetic actions in Weyl semimetals (WSM's) and similar topological matter using standard field theory techniques. Let us  emphasize that  this approach is valid only when the complex many-body system exhibits emergent Lorentz invariance together with  a corresponding spontaneous particle violation within a specific sector of  the Brillouin zone. A crucial distinction here is that now our starting point is not the fundamental fermion-photon interaction in LIV-QED (already corrected  by removing redundant terms), but the many-body Hamiltonian specific to the material being studied. Some of these Hamiltonians can be  derived from first principles, but they commonly emerge from effective approaches or tight-binding approximations. The broad applicability of the SME's fermionic sector, outlined in Eq. (\ref{BASICGENACT}), is beneficial for identifying the precise contributions needed in each instance. This enables us to align the microscopic Hamiltonian of the material  with a Dirac-like theory, effectively mapping it to a relevant sector of the SME. 
In this way  we are left with a  modified Dirac-like action involving fermionic fields $\Psi$'s . However, these fields do not represent physical particles (like electrons and positrons, as in the fundamental interactions approach) but rather emergent quasiparticles that describe the effective fermion degrees of freedom within the material. Their quantization proceeds through standard steps, on top of  the physical vacuum determined by the underlying many-body Hamiltonian, which is experimentally well-defined in the material's rest frame. To obtain effective electromagnetic actions that reveal the material's transport properties, these effective fermion degrees of freedom are subsequently integrated out in a path integral formulation. This integration yields corrections to the standard Maxwell equations, from which corrections to the transport properties are derived.

Let us apply this procedure to our tilted anisotropic two-node WSM to illustrate its functionality. The microscopic Hamiltonian, given by Eq. (\ref{Hamiltonian_Tilting}), is chiral and can be embedded within a $4\times 4$ matrix
\begin{align}
    H= \begin{pmatrix} H_{\chi=-1} & 0 \\  0 & H_{\chi=+1} \end{pmatrix}. 
    \label{INITIALQUIRALHAM}
\end{align}
This Hamiltonian possesses 32 independent parameters, which are contained within $\chi \mbf{b}, \chi b_0, \mbf{v}_\chi$ and $\mathbb A_\chi$. In the chiral representation of the gamma matrices (\ref{CHIRALGAMMA})
the condition $[H, \gamma^5]=0$ then  holds.  
To reformulate the Hamiltonian (\ref{INITIALQUIRALHAM}) as arising from a Dirac-like field theory using the SME input, we must match it with the general expression
\beq
H= \gamma^0 \Gamma^k i \partial_k 
+ \gamma^0 M \eeq
obtained  from the action (\ref{BASICGENACT}). To preserve standard time evolution requires choosing 
$\Gamma^0= \gamma^0$ . While this choice can be achieved via a field redefinition, we directly implement it by setting 
$ c^0{}_\nu= d^0{}_\nu=0$, which imposes 8 constraints on the original parameters. The next step in selecting the appropriate coefficients from  the SME involves demanding   the conditions 
\beq [\gamma^5, \gamma^0 \Gamma^k]=0, \qquad [\gamma^5, \gamma^0 M ]=0,  
\eeq 
implied  by the chirality requirement  $[H, \gamma^5]=0$. These conditions immediately set 
\beq 
e^\mu=f_\mu= g^{\kappa \lambda \mu}= m = H_{\mu\nu}= m_5=0. 
\eeq 
Consequently, the relevant physical parameters in our case are \beq a_\mu, \quad b_\mu, \quad c^\mu{}_\nu \quad {\rm and}\quad d^\mu{}_\nu. 
\label{SMESUF}
\eeq 
These constitute 40 constants. Recalling the 8 constraints already imposed on coefficients  $c^\mu{}_\nu$ and 
$d^\mu{}_\nu$, we are left with precisely 32 parameters, which encode the Hamiltonian of the  two-cone tilted anisotropic WSM in Eq. (\ref{Hamiltonian_Tilting}).
The one-to-one relation among the  set of coefficients (\ref{SMESUF}) and those in the Hamiltonian (\ref{Hamiltonian_Tilting}) is given in the Appendix \ref{APPA}. In this way we have shown  that the SME basis chosen for our purposes  is complete and does not either require or allow for any additional field redefinition in the material rest frame. Going to alternative frames via an observer Lorentz transformation will change the structure of the Hamiltonian and will reshuffle the coefficients among the gamma matrix structure. The study of this possibility is beyond the scope of the present work. 

Summarizing, the  field theory calculation of the  required effective electromagnetic action starts from Eq. (\ref{BASICGENACT}) with the restriction
\beq
\Gamma^\mu = \gamma^\mu + c^\mu{}_\nu \gamma^\nu + d^\mu{}_\nu \gamma^5 \gamma^\nu,
	\qquad
	M = a_\mu \gamma^\mu + b_\mu \gamma^5 \gamma^\mu,
\eeq
where the SME coefficients are fully determined by material parameters set by the band structure (crystal symmetries and velocities) of the material.

 The presence of $\gamma^5$ in the generalized Dirac operator suggests a decomposition into left- and right-handed components, replacing $\gamma^5$ by its eigenvalues $\chi=\pm1$. Following the steps of Ref.~\cite{PhysRevD.109.065005}, we introduce the projectors
\begin{equation}
P_\chi = \frac{1+\chi\gamma^5}{2}, \qquad \chi=\pm1,
\end{equation}
which satisfy
\begin{equation}
P_\chi^2 = P_\chi, \qquad P_+ + P_- = 1, \qquad P_+ P_- = 0,
\end{equation}
and project onto the right-handed ($\chi=+1$) and left-handed ($\chi=-1$) components. Note that $\gamma^\mu P_\chi = P_{-\chi}\gamma^\mu$.

Using these projectors, the Dirac operator decomposes as
\begin{equation}
\Gamma^\mu P_\chi
=
\left(\delta^\mu{}_\nu + c^\mu{}_\nu - \chi d^\mu{}_\nu\right)\gamma^\nu P_\chi
\equiv
\Gamma^\mu_\chi P_\chi,
\end{equation}
which explicitly identifies
\begin{equation}
\Gamma^\mu_\chi = (m_\chi)^\mu{}_\nu \gamma^\nu,
\qquad
(m_\chi)^\mu{}_\nu = \delta^\mu{}_\nu + c^\mu{}_\nu - \chi d^\mu{}_\nu.
\end{equation}

Similarly, the mass-like operator becomes
\begin{equation}
M P_\chi
=
(a_\mu - \chi b_\mu)\gamma^\mu P_\chi
\equiv
(C_\chi)_\mu \gamma^\mu P_\chi,
\end{equation}
with
\begin{equation}
(C_\chi)_\mu = a_\mu - \chi b_\mu.
\end{equation}

In order to obtain a well-defined Hamiltonian description, we impose the condition
\begin{equation}
\Gamma^0 = \gamma^0,
\end{equation}
which ensures that the time derivative retains its standard form and a well-defined Hamiltonian structure. This condition demands
\begin{equation}
c^0{}_\nu = 0, \qquad d^0{}_\nu = 0.
\end{equation}

With these definitions, the fermionic theory reduces to two independent components labeled by $\chi=\pm1$, corresponding to the two Weyl nodes, each characterized by $(m_\chi)^\mu{}_\nu$ and $(C_\chi)_\mu$.

Integrating out the fermions yields the electromagnetic effective action
\begin{align}
\exp\!\big[iS_{\rm eff}(A)\big]
&=\int D\bar{\Psi}D\Psi\;\exp\!\left[i\!\int d^{4}x\,\mathcal L\right]
=\det\!\left(\Gamma^{\mu} i\partial_{\mu}-M-e\,\Gamma^{\mu}A_{\mu}\right),
\label{DET}
\end{align}
which we expand to quadratic order in $A_\mu$:
\begin{equation}
S_{\rm eff}^{(2)}(A)
=\frac{1}{2}\int\!\frac{d^{4}k}{(2\pi)^{4}}\,
A_{\mu}(-k)\;\Pi^{\mu\nu}(k)\;A_{\nu}(k),
\label{DEFL}
\end{equation}
with the vacuum-polarization tensor 
\begin{equation}
\Pi ^{\mu\nu} (k) = -ie^2\\\int \frac{d^4p}{(2\pi)^4}\\\mathrm{tr} [S(p-k)\Gamma^\mu S(p) \Gamma^\nu],
\end{equation}
 computed from the propagator $S(p)=(\Gamma \cdot p-M) ^{-1}$ (evaluated at the external momentum $k$ in practice) \cite{PhysRevD.109.065005}. The linear response current follows as
$J^{\mu}(k)=\delta S_{\rm eff}/\delta A_{\mu}(k)=\Pi^{\mu\nu}(k)\,A_{\nu}(k)$:
its CPT-odd part produces the axion term (schematically $\sim \epsilon^{\mu\nu\alpha\beta}\,\mathcal B_{\alpha}\,A_{\mu}\partial_{\nu}A_{\beta}$, with $\mathcal B_{\alpha}$ fixed by the matched SME parameters and the finite-density state). To stay faithful to the lattice theory, momentum integrals are regularized with the physical hard cutoff $\Lambda$ that delimits the linearization window around each node, ensuring that this field-theoretic treatment and the Berry-curvature semiclassical approach operate within the same regime \cite{Dittrich1985,Dittrich2000}. Up to this point the derivation is independent of whether we deal with type-I or type-II; the distinction enters when extracting transport at finite $\mu$, where Fermi-surface contributions controlled by the tilt become essential.  This setup corresponds to a sub-sector of the minimal QED extension in the SME adapted to solids, with coefficients regarded as band-structure parameters \cite{Colladay1997,Colladay1998}. A closely related construction of CPT-odd effective actions in Lorentz-violating theories exhibiting the axial anomaly was developed in Ref.~\cite{GOMEZ2022137043}, where the structure of the induced axion-like term and its regularization ambiguities were analyzed within a general quantum-field-theoretic framework.

For direct comparison with the semiclassical result, we now write the quadratic response in a form that isolates the same basic integral. The second-order polarization tensor for node $\chi$ is
\begin{equation}
	\Pi^{\mu\nu(2)}_{\chi}
	= i\frac{\delta^j_{\ \rho}}{2}\,\chi e^2\,{(m_\chi^{-1})^\rho}_{\ \lambda}\,
	\epsilon^{\lambda\mu\kappa\nu}\,
	\sum_{s=\pm 1} s\,\mathcal{I'}^{\chi s}_j\, k_\kappa ,
\end{equation}
 and
\begin{equation}\label{Integral2}
	\mathcal{I'}_j^{\chi s}
	= \int \frac{d^{3}\boldsymbol{\mathcal{K}}_\chi}{(2\pi)^3}\,
	\frac{(\mathcal{K}_\chi)_j}{ \|\boldsymbol{\mathcal K}_{\chi}\| ^3}\,
	H \big(\mu - p_{0 s}^{\chi \#}\big),
\end{equation}
with
\begin{align}
	p ^{\chi \#} _{0 s}
	= - p _{j} (m_\chi) ^{j}{}_{0} + (C_\chi) _{0} + s\,\big| {\mathbf{q}} - {\mathbf C} _{\chi} \big|
	= \boldsymbol{\mathcal{V}}_\chi \cdot \boldsymbol{\mathcal{K}}_\chi  + s \, \|\boldsymbol{\mathcal K}_{\chi}\| + E_{\chi 0},
	\label{POLES11}
\end{align}
where $\boldsymbol{\mathcal{K}}_\chi = {\mathbf{q}} - {\mathbf C}_\chi$, $s=\pm1$ is the band index, ${\mathcal{V}}_{\chi}^{\,j}={(m _{\chi}^{-1})^{j}}_{\ i}\,(m _{\chi})^{i}{}_{0}$ is the tilt and $E_{\chi 0}= \boldsymbol{\mathcal V}_{\chi}\!\cdot\!{\mathbf C}_\chi + (C_\chi)_0$. The induced four-current is
\begin{equation}
	J^\mu = \sum_{\chi=\pm1} \Pi^{\mu\nu (2)}_{\chi}\,A_\nu ,
\end{equation}
or, explicitly,
\begin{equation}
	J^{\mu}
	= \sum_{\chi=\pm1}\sum_{s=\pm 1}
	\frac{\delta^j_{\ \rho}}{2}\,s\,\chi\,e^2\,{(m_\chi^{-1})^\rho}_{\ \lambda}\,
	\mathcal{I'}^{\chi s}_j\,\tilde{F}^{\mu \lambda}.
\end{equation}
Using $\tilde{F}^{ij}=\epsilon^{ijk}E^k$ gives the spatial components
\begin{equation}
	J ^{i}
	= \sum_{\chi=\pm1}\sum_{s=\pm 1}
	\frac{s\,\chi}{2}\,e^2\,
	{(m_\chi^{-1})^{l}}_{ j}\,\mathcal{I'}^{\chi s}_l\,
	\epsilon^{ijk}E^k .
\end{equation}
According to Eq. \eqref{relC} , $(C_\chi)_j=\chi\,\tilde{b}^{\,i}(A_\chi)_{ij}$ and the spatial block of $(m_\chi^{-1})$ coincides with the inverse Fermi-velocity matrix of the Weyl cones \eqref{relm},
\begin{equation}
	{(m_\chi^{-1})^{l}}_{j} \;\longrightarrow\; \big(A^{-1}_{\chi}\big)_{lj},
\end{equation}
which encodes anisotropy. With this identification, the integral \eqref{Integral2} matches the semiclassical one in Eq.~\eqref{Integral}, i.e.\ $\mathcal{I'}^{\chi s}_i=\mathcal{I}^{\chi s}_i$, and the induced current density reduces to
\begin{equation}
\label{CurrentDensityF}
	J_{i}
	= \frac{e^{2}}{2 \hbar}\,\epsilon_{i k j}\,E_{k}\,
	\sum_{s=\pm 1}\sum_{\chi=\pm 1}
	s\,\chi\,
	\big(A^{-1}_{\chi}\big)_{lj} \,
	\mathcal{I}_{l}^{\chi s},
\end{equation}
where we have restored $\hbar$ for dimensional consistency (intermediate steps used $\hbar = 1$). Equation \eqref{CurrentDensityF} fully reproduces the chiral-kinetic result \eqref{CurrentDensityFinal} once the same cutoff, tilt, and anisotropy data are used. This establishes the equivalence between the effective-action and Berry-curvature approaches and sets the stage for extracting the anomalous Hall conductivity via the Fermi-sea/Fermi-surface decomposition of $\mathcal{I}^{\chi s}_l$.

\section{Anomalous Hall response in type-II Weyl semimetals} \label{transport_coefficient}

We now focus on the over-tilted (type-II) regime and evaluate the anomalous Hall current by separating, at $T=0$ and finite $\mu$, the Fermi-surface and Fermi-sea contributions implied by the step function in the basic integral. In the type-I case ($\| \boldsymbol{\mathcal{V}} _{\chi} \| < 1$) the Hall response is entirely due to the Fermi-surface piece, which reproduces the axion coefficient; the Fermi-sea piece vanishes after angular integration once the tilt is subcritical. In contrast, in the type-II case ($\| \boldsymbol{\mathcal{V}} _{\chi} \| > 1$) both pieces contribute: the Fermi-surface term captures the pocket geometry, while the Fermi-sea term becomes finite in the cutoff-dependent evaluation and encodes the residual “sea’’ correction. Our starting point is the compact expression for the current density, Eq.~(\ref{CurrentDensityF}), together with the integral definition Eq.~(\ref{Integral}).

To implement the split transparently in $\boldsymbol{\mathcal{I}} _{\chi s}$, we define the band filling (measured from the node) $\Lambda _{\chi} \equiv \mu - E _{\chi 0}$ so that the step function in Eq.~(\ref{Integral}) reads $H \big( \Lambda _{\chi} - \boldsymbol{\mathcal{V}} _{\chi} \cdot \boldsymbol{\mathcal{K}} _{\chi} - s \, \| \boldsymbol{\mathcal{K}} _{\chi} \| \big) $ (cf.~Eq.~\ref{POLES11}). Using $\boldsymbol{\mathcal{K}} _{\chi} / \| \boldsymbol{\mathcal{K}} _{\chi} \| ^{3} = -\, \boldsymbol{\nabla} _{ \boldsymbol{\mathcal{K}} _{\chi}}( \| \boldsymbol{\mathcal{K}} _{\chi} \| ^{-1})$ and integrating by parts, we obtain
\begin{align}
    \boldsymbol{\mathcal I}_{\chi s} = \boldsymbol{\mathcal I}^{\,\text{F-surf}}_{\chi s} + \boldsymbol{\mathcal I}^{\,\text{F-sea}}_{\chi s},
\end{align}
with
\begin{align}
\boldsymbol{\mathcal I}^{\,\text{F-surf}}_{\chi s} & \equiv \int \frac{d ^{3} \boldsymbol{\mathcal{K}} _{\chi} }{(2 \pi ) ^{3}} \, \frac{1}{\| \boldsymbol{\mathcal{K}} _{\chi} \|} \, \boldsymbol{\nabla}_{ \boldsymbol{\mathcal{K}} _{\chi} } H \big( \Lambda _{\chi} - \boldsymbol{\mathcal{V}} _{\chi}  \cdot \boldsymbol{\mathcal{K}} _{\chi} - s \, \| \boldsymbol{\mathcal{K}} _{\chi} \| \big) ,  \label{I_surface} \\[5pt] \boldsymbol{\mathcal I} ^{\,\text{F-sea}} _{\chi s} & \equiv - \int \frac{d ^{3} \boldsymbol{\mathcal{K}} _{\chi} }{( 2 \pi ) ^{3}} \, \boldsymbol{\nabla}_{ \boldsymbol{\mathcal{K}} _{\chi} } \left[\frac{1}{\| \boldsymbol{\mathcal{K}} _{\chi} \|}\, H \big( \Lambda _{\chi} - \boldsymbol{\mathcal{V}} _{\chi} \cdot \boldsymbol{\mathcal{K}} _{\chi} - s \, \| \boldsymbol{\mathcal{K}} _{\chi} \| \big) \right]. \label{I_sea}
\end{align}
Here, $\boldsymbol{\mathcal I} ^{\,\text{F-surf}} _{\chi s}$ localizes on the pocket condition $\Lambda _{\chi} -\boldsymbol{\mathcal{V}} _{\chi}  \cdot \boldsymbol{\mathcal{K}} _{\chi} - s \, \| \boldsymbol{\mathcal{K}} _{\chi} \| = 0$ and provides the Fermi-surface (type-II) contribution that continuously reduces, for $\| \boldsymbol{\mathcal{V}} _{\chi} \| < 1$, to the type-I axion coefficient. The total-divergence structure in \eqref{I_sea} allows a direct use of the divergence theorem, reducing the integral to the outer sphere $\| \boldsymbol{\mathcal{K}} _{\chi} \| = \Lambda$ set by the cutoff; this Fermi-sea (bulk) term vanishes in the subcritical (type-I) regime but becomes finite once the cone is over-tilted, thereby encoding the cutoff-aware sea correction in type-II. We next evaluate the integrals (\ref{I_surface}) and (\ref{I_sea}) in detail.

\subsection{Fermi-surface contribution}

For the pocket contribution \eqref{I_surface} we use the standard rule for the derivative of the Heaviside function,
$\boldsymbol{\nabla}_{ \boldsymbol{\mathcal {K}} _{\chi} } H[g(\boldsymbol{\mathcal {K}} _{\chi})] =\delta [g(\boldsymbol{\mathcal {K}} _{\chi})] \, \boldsymbol{\nabla}_{\boldsymbol{\mathcal {K}} _{\chi}} g (\boldsymbol{\mathcal {K}} _{\chi}) $, for a continuous differentiable function $g (\boldsymbol{\mathcal {K}} _{\chi})$. Using $ g (\boldsymbol{\mathcal {K}} _{\chi}) = \Lambda _{\chi} - \boldsymbol{\mathcal{V}} _{\chi} \cdot \boldsymbol{\mathcal{K}} _{\chi} - s \, \mathcal{K} _{\chi}$, Eq.\eqref{I_surface} yields
\begin{align}
    \boldsymbol{\mathcal{I}} ^{\,\text{F-surf}} _{\chi s} = \int \frac{d ^{3} \boldsymbol{\mathcal{K}} _{\chi} }{( 2 \pi ) ^{3}} \, \frac{1}{\mathcal{K} _{\chi} } \, ( \boldsymbol{\mathcal{V}} _{\chi} + s \, \hat{\boldsymbol{\mathcal K}} _{\chi} ) \; \delta \big( \Lambda _{\chi} - \boldsymbol{\mathcal {V}} _{\chi} \cdot \boldsymbol{\mathcal{K}} _{\chi} - s \, \mathcal{K} _{\chi} \big) , \label{pocket_integral}
\end{align}
i.e. a term localized on the pocket defined by $\Lambda _{\chi} = \boldsymbol{\mathcal {V}} _{\chi} \cdot \boldsymbol{\mathcal{K}} _{\chi} + s \, \mathcal{K} _{\chi} $. By symmetry, the pocket integral in Eq.~\eqref{pocket_integral} can only yield a vector parallel to $\boldsymbol{\mathcal V}_\chi$: after the anisotropy rescaling the measure is isotropic, and the constraint singles out $\boldsymbol{\mathcal V}_\chi$ as the only preferred direction, so transverse components average to zero. We thus write
\begin{equation}
\boldsymbol{\mathcal I} ^{\text{F-surf}} _{s \chi}(\boldsymbol{\mathcal{V}} _{\chi} )  = C _{s} ( \mathcal{V} _{\chi} , \Lambda _{\chi} ) \, \boldsymbol{\mathcal{V}} _{\chi} , 
\end{equation}
with a scalar coefficient $C _{s}$ depending only on the invariants $ ( \mathcal{V} _{\chi} , \Lambda _{\chi} )$. From these expressions we realize that
\begin{align}
C _{s} ( \mathcal{V} _{\chi} , \Lambda _{\chi} ) = \frac{1}{\mathcal{V} ^{2} _{\chi} } \ \int   \, \frac{d ^{3} \boldsymbol{\mathcal{K}} _{\chi}}{(2\pi)^3}  \;    \frac{1}{\mathcal{K} _{\chi} }   \,    (  \mathcal{V} ^{2} _{\chi} + s \boldsymbol{\mathcal{V}} _{\chi} \cdot \hat{ \boldsymbol{\mathcal{K}}} _{\chi} ) \;  \delta \left[ \Lambda _{\chi} -   \mathcal{K} _{\chi} ( s +  \boldsymbol{\mathcal{V}} _{\chi} \cdot \hat{ \boldsymbol{\mathcal{K}}} _{\chi} )  \right] , \label{C_function}
\end{align}
where $\mathcal{V} _{\chi} > 1$ for type-II Weyl semimetals. To evaluate the function \eqref{C_function} we first resolve the constraint with the standard $\delta$-function decomposition:
\begin{align}
    \delta \left[ \Lambda _{\chi} - \mathcal{K} _{\chi} \big( s + \boldsymbol{\mathcal{V}} _{\chi} \cdot \hat{\boldsymbol{\mathcal{K}}} _{\chi} \big) \right] = \frac{ \delta \big( \mathcal{K} _{\chi} - \mathcal{K} ^{\ast} _{s\chi} \big) }{ \big| s + \boldsymbol{\mathcal{V}} _{\chi} \cdot \hat{\boldsymbol{\mathcal{K}}} _{\chi} \big| } \, H \big( \Lambda - \mathcal{K} ^{\ast} _{s\chi} \big) , \qquad \mathcal{K} ^{\ast} _{s\chi} = \frac{\Lambda _{\chi} } { s + \boldsymbol{\mathcal{V}} _{\chi} \cdot \hat{\boldsymbol{\mathcal{K}}} _{\chi} } > 0 , \label{k_radius}
\end{align}
where $\Lambda$ is the hard cutoff that enforces the linearization window. This collapses the radial integral and leaves a purely angular average, with the cutoff entering explicitly through $H ( \Lambda - \mathcal{K} ^{\ast} _{s\chi})$ to discard directions for which the pocket radius would exceed the linear regime. In the over-tilted case, directions with $| \boldsymbol{\mathcal {V}} _{\chi} \cdot \hat{\boldsymbol{\mathcal{K}}} _{\chi}  | > 1$ can yield finite $\mathcal{K} ^{\ast} _{s\chi}$ for both $s = \pm 1$.

Introducing the spherical coordinates $(\mathcal{K} _{\chi} , \theta , \varphi ) $ with the $z$-axis along $\boldsymbol{\mathcal{V}} _{\chi}$, the radial integration can be performed using the Dirac-delta constraint, leaving a purely angular average. With $\boldsymbol{\mathcal{V}} _{\chi} \cdot \hat{\boldsymbol{\mathcal{K}}} _{\chi} = \mathcal{V} _{\chi} \cos \theta$ and a trivial $\varphi$-integral, one arrives at
\begin{align}
C _{ s } ( \mathcal{V} _{\chi} , \Lambda _{\chi} ) =  \frac{ \Lambda _{\chi} }{ 4 \pi ^{2} \mathcal{V}  _{\chi} } \int _{0} ^{\pi}  \;    \frac{ \mathcal{V} _{\chi} + s  \cos \theta }{ ( s  +  \mathcal{V}  _{\chi} \cos \theta ) ^{2} } \; H ( \Lambda -  \mathcal{K} ^{\ast} _{s \chi} ) \, \sin \theta \, d \theta  . 
\end{align}
Changing variables to $x = \cos \theta $, with $ x \in [-1,1]$, we obtain
\begin{align}
C _{ s } ( \mathcal{V} _{\chi} , \Lambda _{\chi} ) =  \frac{ \Lambda _{\chi} }{ 4 \pi ^{2} \mathcal{V} _{\chi} } \int _{-1} ^{+1}  \;    \frac{ \mathcal{V} _{\chi} + s  x }{ ( s  +  \mathcal{V}  _{\chi} x ) ^{2} } \; H ( \Lambda -  \mathcal{K} ^{\ast} _{s \chi} ) \, d x  ,  \label{IntF}
\end{align}
where the cutoff constraint translates to
$0<\dfrac{\Lambda_\chi}{\,s+\mathcal V_\chi x\,}<\Lambda$,
thereby restricting the allowed subinterval of $x\in[-1,1]$ over which the angular integral is performed. 
The resulting $x$-intervals for all sign/tilt combinations, together with the corresponding non-emptiness conditions on $\Lambda$, are summarized in Table~\ref{tab_angular_domains}; the full derivation is relegated to the Appendix \ref{Angular_domain_app} to avoid overloading the main text.

\begin{table}[t]
\centering
\renewcommand{\arraystretch}{2.0}
\begin{tabular}{| c | c | c |}
\hline
$\mathrm{sign}(\Lambda_\chi)$ & Interval for $x$ & Non-emptiness condition on $\Lambda$ \\ 
\hline
$+$ 
& $\displaystyle \big(\max\{-1,\,x^{\ast}_{s\chi}\},\ 1\big)\setminus\{x_{\rm pole}\}$ 
& $\displaystyle \Lambda>\Lambda ^{\ast} _{s \chi +}$ 
\\ \hline
$-$ 
& $\displaystyle \big(-1,\,\min\{1,\,x^{\ast}_{s\chi}\}\big)\setminus\{x_{\rm pole}\}$ 
& $\displaystyle \Lambda> \Lambda ^{\ast} _{s \chi -}$ 
\\
\hline
\end{tabular}
\caption{Angular domains selected by the cutoff for type-II Weyl cones. We define $x ^{\ast} _{s\chi} = (\Lambda _{\chi} / \Lambda - s)/\mathcal{V} _{\chi}$, $x _{\rm pole} = - s/\mathcal{V} _{\chi}$, and $\Lambda ^{\ast} _{s \chi \xi} = \Lambda _{\chi} / ( s + \xi \mathcal{V} _{\chi} )$, with $s = \pm 1$ and $\xi = {\rm sgn} (\Lambda _{\chi} )$. Endpoints are open (strict inequalities), and the pole $x _{\rm pole}$ is excluded if it lies within $(-1,1)$. The non-emptiness conditions ensure that each interval has positive length; in both cases the inequality appears as a lower bound on $\Lambda$ because the right-hand sides $\Lambda ^{\ast} _{s \chi \xi}$ are positive.}
\label{tab_angular_domains}
\end{table}

With the angular restriction in hand, we now evaluate \eqref{IntF} case by case according to Table~\ref{tab_angular_domains}. Performing the integral for fixed $(s,\chi)$ and tilt $\mathcal V _{\chi}$ we obtain
\begin{align}
    \int dx \;    \frac{ \mathcal{V} _{\chi} + s  x }{ ( s  +  \mathcal{V}  _{\chi} x ) ^{2} }  = \frac{ 1 - \mathcal{V} _{\chi} ^{2} }{\mathcal{V} _{\chi} ^{2} }\,\frac{1}{s+\mathcal{V}_{\chi} x} \; + \; \frac{s}{\mathcal{V} _{\chi} ^{2} }\,\ln| s + \mathcal{V} _{\chi} x | \, .
\end{align}


Introducing  the function
\begin{align}
     F _{s \xi } ( \mathcal{V} _{\chi} , \Lambda _{\chi} ) \equiv  \xi   \frac{  \Lambda _{\chi} }{ 4 \pi ^{2} \mathcal{V} _{\chi} ^{3} }  \left\lbrace (s - \xi \mathcal{V} _{\chi}  ) \left[ 1  -  \frac{\Lambda}{\Lambda _{\chi}}  (s + \xi \mathcal{V} _{\chi})  \right] + s \, \ln \Bigg| \frac{\Lambda}{\Lambda _{\chi}}  (s + \xi \mathcal{V} _{\chi})   \Bigg|  \; \right\rbrace,
\end{align}
where $\xi$ denotes the sign of $\Lambda_\chi$,
we obtain  
\begin{align}
C _{ s } ( \mathcal{V} _{\chi} , \Lambda _{\chi} ) =  H( \xi ) \, H (\Lambda - \Lambda ^{\ast} _{s \chi +}) \,  F _{s + } ( \mathcal{V} _{\chi} , \Lambda _{\chi} ) +  H( - \xi ) \, H (\Lambda - \Lambda ^{\ast} _{s \chi -}) \,  F _{s - } ( \mathcal{V} _{\chi} , \Lambda _{\chi} )  , \label{IntF_2}
\end{align}
which simplifies to
\begin{equation}
C _{s}(\mathcal V_\chi,\Lambda_\chi) = H \big(\Lambda-\Lambda^{\ast}_{s\chi \xi}\big) \; F _{s \xi} \big( \mathcal V_\chi,\Lambda_\chi\big) . \label{eq:C-compact}
\end{equation}
The Heaviside factors ensure that the logarithm arguments are positive in the contributing sectors.

Substituting the Fermi-surface contribution 
$\boldsymbol{\mathcal I}^{\,\text{F-surf}}_{\chi s} = C_s(\mathcal V_\chi,\Lambda_\chi)\,\boldsymbol{\mathcal V}_\chi$ 
into Eq.~(\ref{CurrentDensityF}), we obtain the corresponding Hall current,
\begin{align}
    \mathbf J^{\text{F-surf}}
= \frac{e^{2}}{2\hbar} \, \mathbf E \times 
\sum_{s,\chi} s\,\chi\, ( \mathbb A_{\chi}^{-1} ) ^{T} \boldsymbol{\mathcal V}_\chi  \, 
C_s(\mathcal V_\chi,\Lambda_\chi)\,
  .
\label{J_Fsurf_final}
\end{align}
This expression represents the Fermi-surface (pocket) contribution to the anomalous Hall current in the over-tilted regime. 
The coefficient $C_s(\mathcal V_\chi,\Lambda_\chi)$ encodes the cutoff-dependent angular structure of each node, 
ensuring a continuous connection with the type-I limit where the current reduces to the axion form.

\subsection{Fermi-sea contribution}

We now evaluate the bulk (Fermi-sea) piece given by Eq. \eqref{I_sea} with the hard ultraviolet cutoff $\|\boldsymbol{\mathcal K}_\chi\|\le \Lambda$ that delimits the linearization window around each node (cf.\ Sec.~\ref{model}). By the divergence theorem the volume integral reduces to a surface integral on the sphere 
$\|\boldsymbol{\mathcal K}_\chi\|=\Lambda$ with outward unit normal 
$\hat{\boldsymbol{\mathcal K}} _\chi =\boldsymbol{\mathcal K}_\chi/\|\boldsymbol{\mathcal K}_\chi\|$ and $dS=\Lambda^2\,d\Omega$:
\begin{align}
    \boldsymbol{\mathcal I}^{\,\text{F-sea}}_{\chi s}
= -\frac{1}{(2\pi)^3}\oint_{\|\boldsymbol{\mathcal K}_\chi\|=\Lambda}
\frac{1}{\|\boldsymbol{\mathcal K}_\chi\|}\,
H \big(\Lambda_\chi-\boldsymbol{\mathcal V}_\chi \cdot \boldsymbol{\mathcal K}_\chi
- s\,\|\boldsymbol{\mathcal K}_\chi\|\big)\;
\hat{\boldsymbol{\mathcal K}} _\chi \, dS = -\frac{\Lambda}{(2\pi)^3}\int d\Omega\;
H \big[ \Lambda_\chi - \Lambda \, (s+\boldsymbol{\mathcal V}_\chi \cdot \hat{\boldsymbol{\mathcal K}} _\chi ) \big]\;
\hat{\boldsymbol{\mathcal K}} _\chi \,  .
\end{align}
By symmetry the result must be collinear with the tilt $\boldsymbol{\mathcal V}_\chi$, so we choose the polar axis along $\hat{\boldsymbol{\mathcal V}}_\chi = \boldsymbol{\mathcal V}_\chi/\mathcal V_\chi$ and write 
$x = \cos \theta = \hat{\boldsymbol{\mathcal K}} _\chi \cdot \hat{\boldsymbol{\mathcal V}}_\chi $. Then
\begin{align}
    \boldsymbol{\mathcal I}^{\,\text{F-sea}}_{\chi s} = -\frac{\Lambda}{(2\pi) ^{2} } \, \hat{\boldsymbol{\mathcal V}}_\chi  \; \int_{-1}^{1} dx\; x\; H \big[  \Lambda_\chi - \Lambda\,( s+\mathcal V_\chi x) \big]. \label{I_sea_2}
\end{align}
We now proceed in close analogy with the Fermi-sea analysis. The boundary (pocket) contribution inherits the same angular limits and non-emptiness (cutoff) conditions. The only difference lies in the origin of the constraint: for the boundary integral we evaluate on the cutoff sphere $| \boldsymbol{\mathcal K}_\chi|=\Lambda$, so the Heaviside selector is $H \big[ \Lambda_\chi-\Lambda  (\mathcal V_\chi x+s ) \big]$; by contrast, in the sea term the restriction entered through $H \big(\Lambda-\mathcal K^{\ast}_{s\chi}\big)$.

Introducing the angular threshold $x^{\ast}_{s\chi}=(\Lambda_\chi/\Lambda - s)/\mathcal V_\chi$,
equation \eqref{I_sea_2} becomes
\begin{align}
    \boldsymbol{\mathcal I}^{\,\text{F-sea}}_{\chi s} = -\frac{\Lambda}{(2\pi) ^{2} } \, \hat{\boldsymbol{\mathcal V}}_\chi \; \int_{-1}^{1} dx \; x \; H \big[ \mathcal V_\chi\, ( x^{\ast}_{s\chi}-x) \big] . \label{I_sea_3}
\end{align}
The corresponding integration domains for $x$ are summarized in Table~\ref{tab_boundary_selector_checked}, where the intervals are always understood as clipped to $[-1,1]$ and with the pole $x_{\rm pole}$ excluded whenever it lies inside. Their derivation are given in Appendix~\ref{app:boundary_selector_new}.
Using these intervals, the $x$-integral in Eq.~\eqref{I_sea_3} can be rewritten as
\begin{align}
    \boldsymbol{\mathcal I}^{\,\text{F-sea}}_{\chi s} &= -\frac{\Lambda}{4 \pi ^{2} } \, \hat{\boldsymbol{\mathcal V}}_\chi \; \left\lbrace H (\xi ) +  H (- \xi )  \,  H ( \Lambda -  \Lambda ^{\ast} _{s \chi -} )  \right\rbrace \int_{-1}^{\min\{1, x^{\ast}_{s\chi} \}}   x \, dx  . 
\end{align}
Performing these integrals one gets
\begin{align}
    \boldsymbol{\mathcal I}^{\,\text{F-sea}}_{\chi s} &= -\frac{\Lambda}{8 \pi ^{2} } \, \hat{\boldsymbol{\mathcal V}}_\chi \; \left[ H (\xi ) +  H (- \xi )  \,  H ( \Lambda -  \Lambda ^{\ast} _{s \chi -} )  \right] \; \big(\min\{1,x ^{\ast}_{s\chi} \}^{2}-1\big)  .  \label{F-sea_fin}
\end{align}
In the type-I limit ($\mathcal V_\chi <1$) and within the linear window ($|\Lambda_\chi|\ll\Lambda$), the angular threshold 
$x^{\ast}_{s\chi}=(\Lambda_\chi/\Lambda-s)/\mathcal V_\chi$ always lies outside $[-1,1]$ for both $s=\pm1$ (for $\mathcal V_\chi>0$ one has $x^{\ast}_{+\chi}\lesssim -1$ and $x^{\ast}_{-\chi}\gtrsim 1$). Consequently, in the compact expression \eqref{F-sea_fin} the clipped factors reduce to zero in all sign subcases (either $\min\{1,x^{\ast}_{s\chi}\}=\pm1$ or $\max\{-1,x^{\ast}_{s\chi}\}=\pm1$), and the sea term vanishes identically: $\boldsymbol{\mathcal I}^{\,\text{F-sea}}_{\chi s}=\boldsymbol{0} $ $ (\mathcal V_\chi <1,\;|\Lambda_\chi|\ll\Lambda)$. Physically, the Heaviside either selects an empty cap (no contribution) or the full sphere (whose angular moment $\int_{-1}^1 x\,dx=0$ cancels by symmetry).

\begin{table}[t]
\centering
\renewcommand{\arraystretch}{1.8}
\begin{tabular}{| c | c | c |}
\hline
$\mathrm{sign}(\Lambda_\chi)$ & Interval for $x$ & Non-emptiness condition on $\Lambda$ \\ 
\hline
$+$ 
& $\displaystyle \big(-1,\ \min\{1,\,x^{\ast}_{s\chi}\}\big)$ 
& None (always satisfied)
\\ \hline
$-$ 
& $\displaystyle \big(-1,\ \min\{1,\,x^{\ast}_{s\chi}\}\big)$ 
& $\displaystyle \Lambda>\Lambda^{\ast}_{s\chi-} $
\\
\hline
\end{tabular}
\caption{Angular domains selected by the boundary selector 
$H[\mathcal V_\chi(x^{\ast}_{s\chi}-x)]$ 
for $x\in[-1,1]$, $\mathcal V_\chi>1$, and $s=\pm1$, where 
$x^{\ast}_{s\chi}=(\Lambda_\chi/\Lambda - s)/\mathcal V_\chi$ and 
$\Lambda^{\ast}_{s\chi-}=\Lambda_\chi/(s-\mathcal V_\chi)$. 
Because $\mathcal V_\chi>1$, the selector reduces to $H(x^{\ast}-x)$, retaining $x<x^{\ast} _{s\chi}$. 
For $\Lambda_\chi>0$, the inequality $x^{\ast}_{s\chi}>-1$ is always satisfied, so the interval is non-empty for any cutoff. 
For $\Lambda_\chi<0$, the same condition produces the lower cutoff bound $\Lambda>\Lambda^{\ast}_{s\chi-}$.}
\label{tab_boundary_selector_checked}
\end{table}

In contrast, for type-II cones ($\mathcal V_\chi >1$) the threshold $x^{\ast}_{s\chi}$ generically falls inside $[-1,1]$ (subject to the non-emptiness conditions summarized in Table~\ref{tab_boundary_selector_checked}), and the Fermi-sea piece is finite and strictly collinear with the tilt. Using the same compact form one finds
\begin{equation}
\boldsymbol{\mathcal I} ^{\text{F-sea}} _{s \chi}(\boldsymbol{\mathcal{V}} _{\chi} )  = D _{s} ( \mathcal{V} _{\chi} , \Lambda _{\chi} ) \, \boldsymbol{\mathcal{V}} _{\chi} , 
\end{equation}
where
\begin{align}
    D _{s} ( \mathcal{V} _{\chi} , \Lambda _{\chi} ) &=   \frac{\Lambda }{8\pi^{2} \mathcal V_\chi }  \; \left[ 1 - \frac{1}{  \mathcal{V}  _{\chi} ^{2} } \left( \frac{\Lambda _{\chi}}{\Lambda } - s \right) ^{2} \right] \;   \left[ H (\xi ) +  H (- \xi )  \,  H ( \Lambda -  \Lambda ^{\ast} _{s \chi \xi} )  \right] . \label{eq:I-sea-typeII}
\end{align}
Equation~\eqref{eq:I-sea-typeII} implements the angular clipping enforced by the hard cutoff on the sphere $\|\boldsymbol{\mathcal K}_\chi\|=\Lambda$ through the single threshold $x^{\ast}_{s\chi}$: a nontrivial cap is selected whenever $x^{\ast}_{s\chi}\in(-1,1)$ with the appropriate sign of $\mathcal V_\chi$, yielding a nonzero sea contribution that adds to the pocket piece to produce the full anomalous Hall response in the over-tilted regime. Substituting $\boldsymbol{\mathcal I}^{\text{F-sea}}_{s\chi}
= D_s(\mathcal V_\chi,\Lambda_\chi)\,\boldsymbol{\mathcal V}_\chi$
into Eq.~(\ref{CurrentDensityF}) gives
\begin{align}
    \mathbf J^{\text{F-sea}}
= \frac{e^{2}}{2\hbar} \, \mathbf E \times 
\sum_{s,\chi} s\,\chi\, ( \mathbb A_{\chi}^{-1} ) ^{T} \boldsymbol{\mathcal V}_\chi \, 
D_s(\mathcal V_\chi,\Lambda_\chi)\,
  .
\label{J_Fsea_final}
\end{align}
Collecting the two pieces, the anomalous Hall current takes the compact form
\begin{align}
    \mathbf J 
= \frac{e^{2}}{2\hbar} \, \mathbf E \times 
\sum_{s,\chi} s\,\chi\, ( \mathbb A_{\chi}^{-1} ) ^{T} \boldsymbol{\mathcal V}_\chi  \, \left[ C_s(\mathcal V_\chi,\Lambda_\chi) + 
D_s(\mathcal V_\chi,\Lambda_\chi) \, \right]
  .
\label{J_total_final}
\end{align}
This expression makes explicit that the total response is the sum of a Fermi-surface (pocket) part and a cutoff-regulated Fermi-sea (bulk) part. In the type-I regime ($\mathcal V_\chi <1$ and $|\Lambda_\chi|\ll\Lambda$) one finds $D_s=0$, so the Hall response is entirely controlled by $C_s$.  In contrast, in the over-tilted (type-II) regime ($\mathcal V_\chi >1$), both coefficients become finite and additive within the same collinear structure along $\boldsymbol{\mathcal V} _\chi$, while the full tensor anisotropy of the current is encoded in the prefactor $(\mathbb A_\chi^{-1}\boldsymbol{\mathcal V}_\chi)$ appearing in Eq.~\eqref{J_total_final}.

\section{Application} \label{Applications}

To illustrate our formalism, we evaluate the anomalous Hall response in the type-II Weyl semimetal WTe$_2$ under hydrostatic pressure, following the low-energy description introduced in Ref.~\cite{Soluyanov2015_TypeII}. Bulk WTe$_2$ crystallizes in a noncentrosymmetric orthorhombic structure (space group $Pmn2_1$) with lattice parameters $a\simeq3.48~\text{\AA}$, $b\simeq6.28~\text{\AA}$, and $c\simeq14.07~\text{\AA}$. Under a few GPa of pressure, the material realizes a minimal Weyl phase hosting a single pair of strongly tilted Weyl nodes of opposite chirality. In what follows we adopt a node-centered linearized description based on the $k \cdot p$ Hamiltonian extracted from first-principles calculations in Ref.~\cite{Soluyanov2015_TypeII}.

The two Weyl nodes lie in the $k _{z} =0$ plane and are related by inversion of crystal momentum. In our notation they are located at
\begin{align}\label{barb}
\mathbf k_{\chi}=\chi\,\widetilde{\mathbf b},
\qquad
\widetilde{\mathbf b}\simeq(0.015,\,0,\,0)\ \text{\AA}^{-1},
\end{align}
so that the nodes are separated by $2\widetilde{\mathbf b}$ in momentum space and carry opposite chiralities $\chi=\pm1$.  
The energies of the nodes measured from the Fermi level lead to band fillings (measured from each node)
\begin{align}\label{Lambda}
\Lambda_{+}\simeq-0.058~\mathrm{eV}, \qquad
\Lambda_{-}\simeq-0.052~\mathrm{eV}.
\end{align}
Although the Fermi level lies below the nodal energies, in the type-II regime this does not imply that only the valence band contributes: the strong tilt produces directions in momentum space for which the $s=+1$ branch also satisfies $E_{+\chi}<\mu$. Therefore both bands must be retained in the sum $\sum_{s=\pm1}$ of the Berry-curvature current.

The low-energy electronic structure around each Weyl node is described by the linearized Hamiltonian
\begin{align}
H_\chi(\mathbf q)
=
(\tau_\chi q_x+\eta_\chi q_y)\sigma_0
+(\alpha_\chi q_x+\gamma_\chi q_y)\sigma_y
+(\beta_\chi q_x+\delta_\chi q_y)\sigma_z
+\zeta_\chi q_z\,\sigma_x ,
\label{kp_WTe2}
\end{align}
where $\mathbf{q}=\mathbf{k}-\mathbf{k}_\chi$ is the momentum measured from the node, $\sigma_0$ is the identity matrix, and $\boldsymbol{\sigma}=(\sigma_x,\sigma_y,\sigma_z)$ are Pauli matrices in the effective pseudospin space. All coefficients carry units of $\mathrm{eV\,\AA}$ and are obtained from the symmetry-constrained fit to the \textit{ab initio} band structure of Ref. \cite{Soluyanov2015_TypeII}.  
Equation~(\ref{kp_WTe2}) maps directly onto our generic low-energy model
\begin{align}
H_\chi(\mathbf q)=\mathbf v_\chi\!\cdot\!\mathbf q\,\sigma_0
+\mathbf q\,\mathbb A_\chi\,\boldsymbol\sigma ,
\end{align}
with tilt vector $\mathbf v_\chi=(\tau_\chi,\eta_\chi,0)$ and anisotropy matrix (in $\mathrm{eV\,\AA}$ units)
\begin{align}
\mathbb A_\chi=
\begin{pmatrix}
0 & 0 & \zeta_\chi\\
\alpha_\chi & \gamma_\chi & 0\\
\beta_\chi & \delta_\chi & 0
\end{pmatrix} .
\end{align}
For the two nodes the fitted parameters are
\begin{align}
\chi=-1:\quad
\mathbf v_{-}
&=(-2.739,\,0.612,\,0),
&
\mathbb A_{-}
&=
\begin{pmatrix}
0 & 0 & 0.184\\
0.987 & 0 & 0\\
1.107 & 0.270 & 0
\end{pmatrix},
\label{vA-}
\\[6pt]
\chi=+1:\quad
\mathbf v_{+}
&=(1.204,\,0.686,\,0),
&
\mathbb A_{+}
&=
\begin{pmatrix}
0 & 0 & 0.237\\
-1.159 & 0 & 0\\
1.046 & 0.055 & 0
\end{pmatrix},
\label{vA+}
\end{align}
all in units of $\mathrm{eV \, \AA}$. The dimensionless rescaled tilts entering our transport theory are $\boldsymbol{\mathcal{V}} _{\chi} = \mathbb{A} _{\chi} ^{-1} \mathbf{v} _{\chi}$ which are indeed dimensionless since $\mathbb{A} _{\chi} ^{-1}$ has units $(\mathrm{eV \, \AA}) ^{-1}$. Numerically,
\begin{align}
\boldsymbol{\mathcal V}_{-}
&\approx (0.62,\,-2.54,\,-14.9),\qquad
|\boldsymbol{\mathcal V}_{-}|\simeq15.1,\\[5pt]
\boldsymbol{\mathcal V}_{+}
&\approx (-0.59,\,11.3,\,5.08),\qquad
|\boldsymbol{\mathcal V}_{+}|\simeq12.4,
\end{align}
which confirms the strongly overtilted (type-II) character of both cones.

Our integrals are performed in the rescaled momentum $\boldsymbol{\mathcal{K}} _{\chi} = \mathbb{A} _{\chi} \boldsymbol{K} _{\chi}$, which has units of energy (eV).  Consequently the hard cutoff of the linear theory, $\|\boldsymbol{\mathcal K}_\chi\|\le\Lambda$, must also be specified in energy units.   Starting from a conservative momentum window $\|\boldsymbol{K} _{\chi} \|\lesssim \Lambda _{k}$ with $\Lambda _{k} = 0.2~\text{\AA} ^{-1}$, we map it to $\boldsymbol{\mathcal{K}}$-space through the characteristic velocity (in $\mathrm{eV\,\AA}$ units)
\begin{align}
\bar v_\chi = |\det(\mathbb A_\chi)|^{1/3}  ,
\end{align}
and define the node-dependent cutoffs (in \text{eV} units)
\begin{align}
\Lambda_\chi^{(\mathrm{cut})}=\bar v_\chi\,\Lambda_k ,
\end{align}
yielding
\begin{align}
\Lambda _{-} ^{(\mathrm{cut})} \simeq 0.073~\mathrm{eV},\qquad
\Lambda _{+} ^{(\mathrm{cut})} \simeq 0.049~\mathrm{eV}.
\end{align}
The anomalous Hall current follows from our general result
\begin{align}
\mathbf J
=
\frac{e^{2}}{2\hbar}\,
\mathbf E\times
\sum_{\chi=\pm1}\sum_{s=\pm1}
s\,\chi\,
  { (\mathbb A_\chi^{-1} ) ^{T}} \boldsymbol{\mathcal V}_\chi \,
\Big[ C_s(\mathcal V_\chi,\Lambda_\chi)+D_s(\mathcal V_\chi,\Lambda_\chi)\Big],
\label{J_app}
\end{align}
where $C _{s}$ and $D _{s}$ are the Fermi-surface and Fermi-sea coefficients of Eqs.~(\ref{eq:C-compact}) and (\ref{eq:I-sea-typeII}).  
In Eq.~(\ref{J_app}) the vector $\mathbb A_\chi^{-1}\boldsymbol{\mathcal V}_\chi$ has units $(\mathrm{eV\,\AA})^{-1}$, while $C _{s}$ and $D _{s}$ have units of eV; therefore their product carries units $\text{\AA}^{-1}$, as required for a three-dimensional Hall response.

For WTe$_{2}$ we have $\Lambda _{\chi} < 0$ for both nodes, hence $\xi = \mathrm{sgn}(\Lambda _{\chi}) = - 1$. Using the cutoffs above, the coefficients (in eV) are
\begin{align}
\chi=-1:\quad
&C_{-1}\simeq-1.18\times10^{-4},\quad
C_{+1}\simeq-1.15\times10^{-4},\\
&D_{-1}\simeq+6.14\times10^{-5},\quad
D_{+1}\simeq+6.06\times10^{-5},\\[4pt]
\chi=+1:\quad
&C_{-1}\simeq-9.34\times10^{-5},\quad
C_{+1}\simeq-8.83\times10^{-5},\\
&D_{-1}\simeq+5.05\times10^{-5},\quad
D_{+1}\simeq+4.89\times10^{-5}.
\end{align}
These values make explicit that in the type-II regime both $s=\pm1$ contribute with comparable magnitude; retaining only $s=-1$ would miss a substantial part of the response.

Defining the geometric vectors (in $(\mathrm{eV\,\AA})^{-1}$)
\begin{align}
\mathbf W_\chi\equiv  { (\mathbb A_\chi^{-1} ) ^{T}} \boldsymbol{\mathcal V}_\chi
,
\end{align}
we obtain, for the two nodes,
\begin{align}
\mathbf W_{-}\simeq(-80.90,\;11.19,\;-9.42),\qquad
\mathbf W_{+}\simeq(21.44,\;185.22,\;204.67),
\end{align}
all in units of $(\mathrm{eV\,\AA})^{-1}$. Equation~(\ref{J_app}) can then be written in the standard form
\begin{align}
\mathbf J
=
\frac{e^{2}}{h}\,
\mathbf E\times \boldsymbol{\mathcal G},
\qquad
\boldsymbol{\mathcal G}
=\pi\sum_{\chi=\pm1}\sum_{s=\pm1}
s\,\chi\,\mathbf W_\chi\,
\big[C_s(\mathcal V_\chi,\Lambda_\chi)+D_s(\mathcal V_\chi,\Lambda_\chi)\big],
\label{G_def_app}
\end{align}
with $\boldsymbol{\mathcal G}$ in $\text{\AA}^{-1}$. Using the numerical coefficients quoted above (in eV) for $C_s$ and $D_s$, we find
\begin{align}
\boldsymbol{\mathcal G}
\simeq
(7.95\times10^{-4},\;
1.96\times10^{-3},\;
2.32\times10^{-3})\ \text{\AA}^{-1}.
\label{G_num_app}
\end{align}
The anomalous Hall conductivity tensor follows as
\begin{align}
\sigma ^{\mathrm{A}}_{ij}
=
\frac{e^{2}}{h}\,
\varepsilon_{ijk}\,\mathcal G_{k},
\end{align}
or explicitly ($\text{entries in }\text{\AA}^{-1}$)
\begin{align}
\boldsymbol{\sigma}^{\mathrm A}
\simeq
\frac{e^{2}}{h}
\begin{pmatrix}
0 & 2.32\times10^{-3} & -1.96\times10^{-3}\\
-2.32\times10^{-3} & 0 & 7.95\times10^{-4}\\
1.96\times10^{-3} & -7.95\times10^{-4} & 0
\end{pmatrix} .
\label{sigma_tensor_app_new}
\end{align}
Then we obtain
\begin{align}
\sigma^{\mathrm A}_{xz}&\simeq-7.59~\text{S/cm},\qquad
\sigma^{\mathrm A}_{xy}\simeq+8.99~\text{S/cm},\qquad
\sigma^{\mathrm A}_{yz}\simeq+3.08~\text{S/cm}.
\label{sigma_3D_app_new}
\end{align}
We stress that bulk WTe$_2$ preserves time-reversal symmetry at zero magnetic field, so a spontaneous anomalous Hall effect is not symmetry protected and can be partially or fully canceled once all symmetry-related Weyl nodes are included. Nevertheless, the tensor above represents the intrinsic contribution associated with a single strongly tilted minimal pair and provides a natural scale for Hall-like responses observed in magnetotransport experiments, such as the planar Hall effect and related anisotropic signals reported in WTe$_2$~\cite{Wang2018_PlanarHall,Luo2015_WTe2Hall}.  

The magnitude of the predicted anomalous Hall conductivity is smaller than the longitudinal conductivity typically measured in this material, yet it remains sizable compared with experimentally accessible Hall signals~\cite{Luo2015_WTe2Hall,Zhou2016_WTe2PRB}. This highlights the important role played by both the Fermi-surface and the cutoff-regulated Fermi-sea contributions in type-II Weyl semimetals, and underscores the strong sensitivity of the Hall response to the tilt and anisotropy of the Weyl cones~\cite{Zyuzin2016_AHE_TypeII}. The present calculation also demonstrates explicitly that both bands $s=\pm1$ must be retained in the overtilted regime in order to obtain a consistent anomalous Hall response within the linearized theory.

\section{Discussion} \label{conclusion}

In this work we have developed a unified and cutoff-dependent description of the CPT-odd electromagnetic response of tilted, anisotropic Weyl semimetals in the overtilted (type-II) regime. By combining a Berry-curvature-based chiral kinetic theory with a field-theoretic derivation of the effective action, we have shown that the anomalous Hall response at finite density naturally decomposes into Fermi-surface and Fermi-sea contributions, both of which become essential once the Weyl cones are overtilted. This extends the axion electrodynamics paradigm beyond the ideal type-I limit and clarifies how Lorentz-violating band-structure effects renormalize and supplement the axion-like response.

A key result is that, while the axion-like structure of the CPT-odd response survives across the type-I to type-II Lifshitz transition, its coefficient acquires explicit dependence on the tilt magnitude, anisotropy, and ultraviolet cutoff. In the type-I regime the Fermi-sea contribution vanishes within the linearized theory, and the Hall response is entirely controlled by the Fermi-surface term, reproducing the familiar axionic result. In contrast, for type-II cones the presence of open electron and hole pockets activates a finite Fermi-sea contribution that is intrinsically cutoff dependent and collinear with the tilt direction. Physically, this term encodes the residual contribution of filled states at the boundary of the linearization window, which cannot be discarded once the dispersion becomes unbounded.

Our explicit evaluation for pressurized WTe$_2$ illustrates this mechanism in a realistic setting. Using parameters extracted from first-principles calculations and experiments, we find that Fermi-surface and Fermi-sea contributions yield a finite but  anisotropic anomalous Hall conductivity. The dominant tensor components reflect the combined effect of strong tilt and pronounced velocity anisotropy, providing a clear transport signature of the type-II regime. While bulk WTe$_2$ preserves time-reversal symmetry and therefore does not exhibit a symmetry-protected spontaneous anomalous Hall effect at zero magnetic field, the intrinsic tensor derived here sets a natural scale for Hall-like responses observed in magnetotransport, such as the planar Hall effect and related anisotropic signals.

To make direct contact with the SME we provided the one to one correspondence between the parameters that characterize the microscopic Hamiltonian of a  two-cone tilted anisotropic Weyl semimetal and those we have selected from the fermionic sector of the SME. These identifications  show that our selection of basis in the SME is unique for this system. For the particular case of $\mathrm{WTe}_2$ the SME coefficients are explicitly calculated. Not surprisingly  the describe a huge emergent Lorentz symmetry violation, which is expected in matter. For example, the coefficients $a_0 $ and $b_0 $ are of the order of  $\mathrm{meV}$'s,
which compare favorably with the the Fermi energy $\mu \equiv 55 \, \mathrm{meV}$,  while $a_j$ and $b_j$ are  suppressed only by one order of magnitude. Also, many components of the dimensionless  coefficients $c^i{}_j$ and  $d^i{}_j$ are of order one.  These values are specific to the condensed-matter realization considered here and are fixed by the band-structure parameters of the material in its rest frame. In contrast, in the high-energy physics context, SME coefficients are constrained by high-precision experiments to be extremely small. For example, roughly speaking the  coefficients  $c^i{}_j$,   $d^i{}_j$  are typically bounded in a range $10^{-16}- 10^{-14}$.  \cite{Kostelecky:2008ts}.

From a broader perspective, our results highlight the necessity of a physical ultraviolet regularization when addressing electromagnetic response in type-II Weyl systems. The hard momentum cutoff tied to the lattice bandwidth provides a transparent and physically motivated prescription that reconciles the semiclassical and field-theoretic approaches. This cutoff sensitivity is not an artifact, but rather a genuine feature of overtilted dispersions, where the linear theory alone is insufficient to fully capture the response without reference to the underlying lattice.

Several extensions of the present work are natural. On the theoretical side, it would be interesting to incorporate smooth or anisotropic cutoff schemes derived from explicit tight-binding models, and to explore how disorder and finite temperature modify the balance between Fermi-surface and Fermi-sea contributions. Generalizations to frequency-dependent response functions could also address optical Hall and magneto-electric effects in the type-II regime. On the materials side, the framework developed here can be readily applied to other candidate type-II Weyl semimetals such as MoTe$_2$, TaIrTe$_4$, or noncentrosymmetric Weyl systems with multiple node pairs and reduced symmetry, where cancellations between different nodes may be incomplete. Finally, extending the analysis to driven or nonequilibrium settings may shed light on how CPT-odd responses manifest under strong fields or in ultrafast experiments.

Overall, our work establishes a consistent and physically grounded description of CPT-odd electrodynamics in Weyl semimetals that continuously connects the ideal axionic response of type-I systems to the strongly Lorentz-violating type-II regime, and provides a solid basis for interpreting anomalous Hall phenomena in realistic materials.

\acknowledgments{R.M.v.D., A.M.-R., and L.F.U. acknowledge financial support by UNAM-PAPIIT project No. IG100224, UNAM-PAPIME project No. PE109226 and by SECIHTI project No. CBF-2025-I-1862. A.M.-R. gratefully acknowledges additional support from the Marcos Moshinsky Foundation. R.M.v.D. was supported by UNAM Posdoctoral Program (POSDOC).}

\appendix

\section{Matching between SME and condensed-matter parameters} \label{match}

\label{APPA}

We start from the Dirac-like Lagrangian \eqref{BASICGENACT} in natural units ($c\rightarrow v_f=\hbar  = 1$),
\begin{equation}
	\mathcal{L} = \bar{\Psi}\left( \Gamma^\mu i\partial_\mu - M \right)\Psi,
\end{equation}
with
\begin{equation}
	\Gamma^\mu = \gamma^\mu + c^\mu{}_\nu \gamma^\nu + d^\mu{}_\nu \gamma^5 \gamma^\nu,
	\qquad
	M = a_\mu \gamma^\mu + b_\mu \gamma^5 \gamma^\mu.
\end{equation}
The replacement $c \to v_f$ reflects that the effective theory describes quasiparticles in a medium, where the characteristic velocity is given by the Fermi velocity rather than the speed of light.

Imposing $\Gamma^0 = \gamma^0$, which requires $c^0{}_\nu = 0$ and $d^0{}_\nu = 0$, 
the Hamiltonian in momentum space reads
\begin{equation}
	\mathcal{H} = -\gamma^0 \Gamma^i k_i + \gamma^0 M
\end{equation}
In units $\hbar=1= v_f$ the wave vector $k_i$ and  $M$ have dimensions of $\text eV$. To establish the correspondence with the condensed-matter Hamiltonian, dimensional factors are restored at this stage.  The Fermi velocity $v_f$ is introduced together with $\hbar$ to ensure that the kinetic term has the correct energy dimension. Then, the Hamiltonian takes the form
\begin{equation}
	H = -\hbar v_f\, \gamma^0 \Gamma^i k_i + \gamma^0 M,
	\label{hamU}
\end{equation}
where the dimension of $k_i$ is now ${\AA^{-1}}$ and $\Gamma^i$ are still dimensionless. The operator $M$ is left unchanged, as it already carries the appropriate energy dimension and encodes material-dependent parameters without requiring additional rescaling.

In the chiral representation of the Dirac matrices,
\begin{equation}
	\gamma^0 =
	\begin{pmatrix}
		0 & \sigma_0 \\
		\sigma_0 & 0
	\end{pmatrix},
	\qquad
	\gamma^i =
	\begin{pmatrix}
		0 & \sigma^i \\
		-\sigma^i & 0
	\end{pmatrix},
	\qquad
	\gamma^5 =
	\begin{pmatrix}
		-\sigma_0 & 0 \\
		0 & \sigma_0
	\end{pmatrix},
    \label{CHIRALGAMMA}
\end{equation}
the Hamiltonian separates into left- and right-handed contributions.

We introduce the parametrization
\begin{equation}
	(m_\chi)^\mu{}_\nu = \delta^\mu{}_\nu + c^\mu{}_\nu - \chi d^\mu{}_\nu
	=
	\begin{pmatrix}
		1 & 0 \\
		V_\chi^i & (U_\chi)^i{}_j
	\end{pmatrix},
\end{equation}
and
\begin{equation}
(C_\chi)_\mu = a_\mu - \chi b_\mu.
\end{equation}
A direct calculation yields
\begin{equation}
	H_L
	=
	- \hbar v_f V_L^i k_i + \hbar v_f k_i (U_L)^i{}_j \sigma^j + (C_L)_0 - (C_L)_j \sigma^j,
\end{equation}
\begin{equation}
	H_R
	=
	- \hbar v_f V_R^i k_i - \hbar v_f k_i (U_R)^i{}_j \sigma^j + (C_R)_0 + (C_R)_j \sigma^j.
\end{equation}

These expressions are compared with the low-energy Hamiltonian of a Weyl semimetal \eqref{Hamiltonian_Tilting}
\begin{equation}
	H_\chi(\mathbf k)
	=
	\mathbf v_\chi \cdot (\mathbf k + \chi \tilde{\mathbf b})
	-
	\chi \tilde b_0 \sigma_0
	+
	\chi (\mathbf k + \chi \tilde{\mathbf b})_i (A_\chi)_{ij} \sigma^j.
\end{equation}
In square bracket for the inside variable we recall the units in which the matter parameters are presented: $[H]=\text{eV} , \,\, [{\mathbf v}_\chi] = \text{eV\,\AA}, \,\, [{\mathbf k}]= \text{\AA} ^{-1}, \,\, [ \tilde{\mathbf b}]= \text{\AA} ^{-1}, \,\, [\tilde b_0]=\text{eV}, \,\,  [\mathbb{A}_\chi]= \text{eV\,\AA}$ 

For $\chi = +1, -1$ (R, L, respectively), one obtains
\begin{equation}
	H_L
	=
	\mathbf v_L \cdot \mathbf k
	-
	\mathbf v_L \cdot \tilde{\mathbf b}
	+
	\tilde b_0
	-
	k_i (A_L)_{ij} \sigma^j
	+
	\tilde b_i (A_L)_{ij} \sigma^j,
\end{equation}
\begin{equation}
	H_R
	=
	\mathbf v_R \cdot \mathbf k
	+
	\mathbf v_R \cdot \tilde{\mathbf b}
	-
	\tilde b_0
	+
	k_i (A_R)_{ij} \sigma^j
	+
	\tilde b_i (A_R)_{ij} \sigma^j.
\end{equation}
Therefore, matching the coefficients, we identify
\begin{equation}\label{relm}
	(m_\chi)^i{}_0 = V_\chi^i = \frac{1}{\hbar v_f} v_\chi^i,
	\qquad
	(m_\chi)^i{}_j = (U_\chi)^i{}_j = \frac{1}{\hbar v_f} (A_\chi)_{ij},
\end{equation}
\begin{equation}\label{relC}
	(C_\chi)_0 = \chi\left(v_\chi^i \tilde b_i - \tilde b_0\right),
	\qquad
	(C_\chi)_j = \chi\, \tilde b_i (A_\chi)_{ij},
\end{equation}
\begin{equation}
	a_0 = \frac{1}{2}\,\tilde b_i (v_R^i - v_L^i),
	\qquad
	b_0 = \tilde b_0 - \frac{1}{2}\,\tilde b_i (v_R^i + v_L^i),
\end{equation}
\begin{equation}
	a_j = \frac{1}{2}\,\tilde b_i \big[(A_R)_{ij} - (A_L)_{ij}\big],
	\qquad
	b_j = -\frac{1}{2}\,\tilde b_i \big[(A_R)_{ij} + (A_L)_{ij}\big].
\end{equation}

From the definition of $(m_\chi)^\mu{}_\nu$, we finally obtain
\begin{equation}
	c^i{}_0 = \frac{1}{\hbar v_f}\frac{1}{2}(v_R^i + v_L^i),
	\qquad
	d^i{}_0 = \frac{1}{\hbar v_f}\frac{1}{2}(v_L^i - v_R^i),
\end{equation}
\begin{equation}
	c^i{}_j = \frac{1}{\hbar v_f}\frac{1}{2} \big[(A_R)_{ij} + (A_L)_{ij}\big] - \delta^i{}_j,
	\qquad
	d^i{}_j = \frac{1}{\hbar v_f}\frac{1}{2}\big[(A_L)_{ij} - (A_R)_{ij}\big].
\end{equation}

Using the parameters of the type-II Weyl semimetal WTe$_2$, namely $\tilde{\mathbf b}$ from Eq.\eqref{barb}, $\tilde b_0$ extracted from the node energies $\Lambda_\pm$ in Eq.\eqref{Lambda} ($\Lambda_\pm =\mu \pm {\tilde b}_0,  \, \mu= -0.055\,\text{eV} $), and the velocities $\mathbf{v}_\chi$ and anisotropy matrices $\mathbb{A}
_\chi$ from Eqs.\eqref{vA-} and \eqref{vA+}, we obtain the following numerical values for the SME parameters ($v_f= 1\times 10^6\ \text{m/s}, \implies \, \hbar v_f \approx 6.6  \,\,  \text{eV\,\AA },$)
\begin{equation}
	a_0 \approx 0.0296\ \mathrm{eV}, \qquad
	b_0 \approx 0.0085\ \mathrm{eV},
\end{equation}
\begin{equation}
	a_j \approx (0,\,0,\,0.0004)\ \mathrm{eV}, \qquad
	b_j \approx (0,\,0,\,-0.0032)\ \mathrm{eV},
\end{equation}
\begin{equation}
	c^i{}_0 \approx (-0.1163,\,0.0983,\,0), \qquad
	d^i{}_0 \approx (-0.2897,\,-0.0056,\,0),
\end{equation}
\begin{equation}
	c^i{}_j \approx
	\begin{pmatrix}
		-1 & 0 & 0.0319 \\
		-0.0130 & -1 & 0 \\
		0.1631 & 0.0246 & -1
	\end{pmatrix},
	\quad
	d^i{}_j \approx
	\begin{pmatrix}
		0 & 0 & -0.0040 \\
		0.1625 & 0 & 0 \\
		0.0046 & 0.0163 & 0
	\end{pmatrix}.
\end{equation}

These values are specific to the condensed-matter realization considered here and are fixed by the band-structure parameters of the material. In contrast, in the high-energy physics context, SME coefficients are constrained by high-precision experiments to be extremely small \cite{Kostelecky:2008ts}.

\section{Angular domain selected by the cutoff}
\label{Angular_domain_app}

For type-II Weyl cones ($\mathcal V_\chi>1$), the cutoff condition
\begin{align}
0 < \frac{\Lambda_\chi}{ s + \mathcal V _{\chi} x } < \Lambda , \qquad s = \pm 1 , \quad \Lambda > 0 ,  \label{app:ineq-Vpos}
\end{align}
selects a restricted angular range $x = \cos \theta \in [-1,1]$ that defines the portion of momentum space contributing to Eq.~\eqref{IntF}.
We introduce the shorthand
\begin{align}
x ^{\ast} _{s\chi} = \frac{\Lambda _{\chi} / \Lambda - s}{\mathcal{V} _{\chi} } , \qquad x _{\rm pole} = -\frac{s}{\mathcal{V} _{\chi}} ,  \qquad \Lambda ^{\ast} _{s\chi \pm } = \frac{\Lambda _{\chi} }{ s \pm \mathcal{V} _{\chi} } ,
\end{align}
where $x ^{\ast} _{s\chi}$ locates the boundary set by the cutoff, $x _{\rm pole}$ denotes the angular position of the pole of the integrand [excluded if it lies in $(-1,1)$], and $\Lambda ^{\ast} _{s\chi \pm }$ defines the minimal cutoff amplitude required for a nonempty integration domain.

\subsection*{Case $\Lambda _{\chi} > 0$}

The inequality~\eqref{app:ineq-Vpos} becomes
\begin{align}
s + \mathcal{V} _{\chi} x > \frac{\Lambda _{\chi} }{\Lambda} ,
\end{align}
which yields
\begin{align}
x > x ^{\ast} _{s\chi}.
\end{align}
Intersecting with $x \in (-1,1)$ gives the allowed interval
\begin{align}
x \in \big( \max\{-1,\,x^{\ast}_{s\chi}\},\,1\big)\setminus\{x_{\rm pole}\} . 
\end{align}
Non-emptiness ($x ^{\ast} _{s\chi} <1$) is equivalent to the cutoff condition
\begin{align}
\frac{\Lambda _{\chi}/ \Lambda - s}{\mathcal{V} _{\chi}} < 1 \quad \Longleftrightarrow \quad \frac{\Lambda _{\chi} }{\Lambda} < s + \mathcal{V} _{\chi} \quad \Longleftrightarrow \quad \Lambda > \Lambda ^{\ast} _{s\chi+} . \label{app:cutoff-Vpos}
\end{align}

\subsection*{Case $\Lambda_\chi<0$}

For negative band filling, Eq.~\eqref{app:ineq-Vpos} reads
\begin{align}
s + \mathcal{V} _{\chi} x < \frac{\Lambda _{\chi} }{\Lambda},
\end{align}
and the allowed region is now
\begin{align}
x < x ^{\ast} _{s\chi},
\end{align}
namely
\begin{align}
x \in \big( -1,\,\min\{1,\,x^{\ast}_{s\chi}\}\big)\setminus\{x_{\rm pole}\},
\end{align}
which exists when
\begin{align}
\frac{\Lambda_\chi}{\Lambda}>s-\mathcal V_\chi
\quad\Longleftrightarrow\quad
\Lambda>\Lambda^{\ast}_{s\chi-} . 
\end{align}

\

These two cases ($\Lambda_\chi\gtrless0$) fully describe the integration domain for the type-II regime ($\mathcal V_\chi>1$).
The corresponding angular ranges and cutoff thresholds are summarized in Table~\ref{tab_angular_domains}.

\section{Angular domains from the boundary selector}
\label{app:boundary_selector_new}

In this appendix we determine the polar angles $x=\cos\theta\in[-1,1]$ (endpoints have zero measure) selected by the boundary Heaviside selector
\begin{equation}
H \big[\mathcal V_\chi\,(x^{\ast}_{s\chi}-x)\big],
\qquad s=\pm1,\qquad \Lambda>0,
\label{appB:selector}
\end{equation}
working throughout in the overtilted regime $\mathcal V_\chi>1$. 
Since the selection in Eq.~\eqref{appB:selector} is strict, the endpoint $x=x^{\ast}_{s\chi}$ is open, and the pole 
$x_{\rm pole}=-s/\mathcal V_\chi$, if it lies in $(-1,1)$, must be excluded.

The boundary selector~\eqref{appB:selector} differs conceptually from the cutoff condition discussed in 
Appendix~\ref{Angular_domain_app}, where the admissible angular domain followed from the double inequality (\ref{app:ineq-Vpos}). That condition simultaneously enforces the sign of the denominator and the upper cutoff~$\Lambda$, 
so the resulting intervals depended on both $\mathrm{sign}(\Lambda_\chi)$ and the tilt amplitude, 
and their non-emptiness required explicit lower bounds on~$\Lambda$. 

By contrast, the selector~\eqref{appB:selector} plays a purely geometric role: 
it identifies the side of the boundary $x=x^{\ast}_{s\chi}$ to be retained, 
without reference to the cutoff scale or to the positivity of $\Lambda_\chi/(s+\mathcal V_\chi x)$.  
Under $\mathcal V_\chi>1$, the argument of the Heaviside is positive for $x<x^{\ast}_{s\chi}$, 
so the selector simply keeps that region:
\begin{equation}
x<x^{\ast}_{s\chi}.
\label{appB:side}
\end{equation}
The integration interval is therefore set by $x^{\ast}_{s\chi}$ and the sphere limits, 
excluding $x_{\rm pole}$ if it lies within~$(-1,1)$.

\subsection*{Case $\Lambda_\chi>0$}

For positive band filling, Eq.~\eqref{appB:side} selects
\begin{equation}
x\in\big(-1,\ \min\{1,\,x^{\ast}_{s\chi}\}\big)\setminus\{x_{\rm pole}\},
\label{appB:interval-posL}
\end{equation}
which is non-empty provided $x^{\ast}_{s\chi}>-1$, i.e.
\[
\frac{\Lambda_\chi}{\Lambda}>s-\mathcal V_\chi.
\]
Since $\mathcal V_\chi>1$ and $\Lambda_\chi/\Lambda>0$, this inequality holds automatically for any $\Lambda>0$.  
Thus, for $\Lambda_\chi>0$ the selected angular domain is always non-empty and independent of the cutoff.

\subsection*{Case $\Lambda_\chi<0$}

When the band filling is negative, the geometric interval \eqref{appB:interval-posL} remains valid, 
but the sign reversal of $\Lambda_\chi$ makes the non-emptiness condition nontrivial.  
Requiring $x^{\ast}_{s\chi}>-1$ gives
\begin{equation}
\frac{\Lambda_\chi}{\Lambda}>s-\mathcal V_\chi
\quad\Longleftrightarrow\quad
\Lambda>\Lambda^{\ast}_{s\chi-},
\qquad
\Lambda^{\ast}_{s\chi-}=\frac{\Lambda_\chi}{s-\mathcal V_\chi}.
\label{appB:cutoff-negL}
\end{equation}
Because both $\Lambda_\chi<0$ and $s-\mathcal V_\chi<0$, the right-hand side is positive, 
so the condition defines a finite lower bound on $\Lambda$.  
Hence, for $\Lambda_\chi<0$, the boundary selector reproduces the same cutoff threshold 
$\Lambda>\Lambda^{\ast}_{s\chi-}$ obtained in Appendix~\ref{Angular_domain_app}.

\bibliography{Referencias.bib}

\end{document}